\documentclass[journal=inocaj,manuscript=article,layout=twocolumn,fontsize=10pt  
]{achemso}
\usepackage{achemso}
\SectionNumbersOn
\usepackage[version=4]{mhchem} 

\usepackage{siunitx} 
\usepackage{amssymb}
\usepackage{amsfonts}
\usepackage{amsmath}
\usepackage{graphicx}
\usepackage{graphicx, array, blindtext}
\usepackage{float}
\usepackage{color}
\usepackage{longtable}
\usepackage{revsymb}
\usepackage{units}
\usepackage{nicefrac}
\usepackage{textcomp}
\usepackage{epsfig}
\usepackage{footmisc}
\usepackage{soul}%
\usepackage{comment}
\usepackage{adjustbox}
\usepackage{booktabs,eqparbox}
\newcommand{\onlinecite}[1]{\hspace{-1 ex} \nocite{#1}\citenum{#1}}
\usepackage{multirow}  
\title{Structural metastability and Fermi surface Topology of SrAl$_2$Si$_2$}
\author{Stamatios Strikos}
\email{stamostrik@pos.if.ufrj.br}
\affiliation{Instituto de Física, Universidade Federal do Rio de Janeiro, CxP 68528, 21945-972 Rio de Janeiro, Brazil}
\author{Boby Joseph}
\affiliation{Elettra-Sincrotrone Trieste, S.S. 14 – km 163,5, Area Science Park 34149 Basovizza, Trieste, Italy}
\author{Frederico G. Alabarse}
\affiliation{Elettra-Sincrotrone Trieste, S.S. 14 – km 163,5, Area Science Park 34149 Basovizza, Trieste, Italy}
\author{George Valadares}
\affiliation{Universidade Federal do Acre, Rodovia BR 364, Km 04, Distrito Industrial, Rio Branco, AC, 69920-900, Brazil}
\author{Deyse G. Costa}
\affiliation{Departamento de Química, Universidade Federal da Viçosa, Viçosa, Caixa Postal 216, Brazil}
\author{Rodrigo B. Capaz}
\affiliation{Instituto de Física, Universidade Federal do Rio de Janeiro, CxP 68528, 21945-972 Rio de Janeiro, Brazil}
\author{Mohammed ElMassalami}
\email{massalam@if.ufrj.br}
\affiliation{Instituto de Física, Universidade Federal do Rio de Janeiro, CxP 68528, 21945-972 Rio de Janeiro, Brazil}
\begin{document}
\maketitle
\begin{tocentry}
\begin{center}
\includegraphics[width=1.0\columnwidth,trim={0.2cm 0.2cm 0.2cm 0.15cm}, clip]{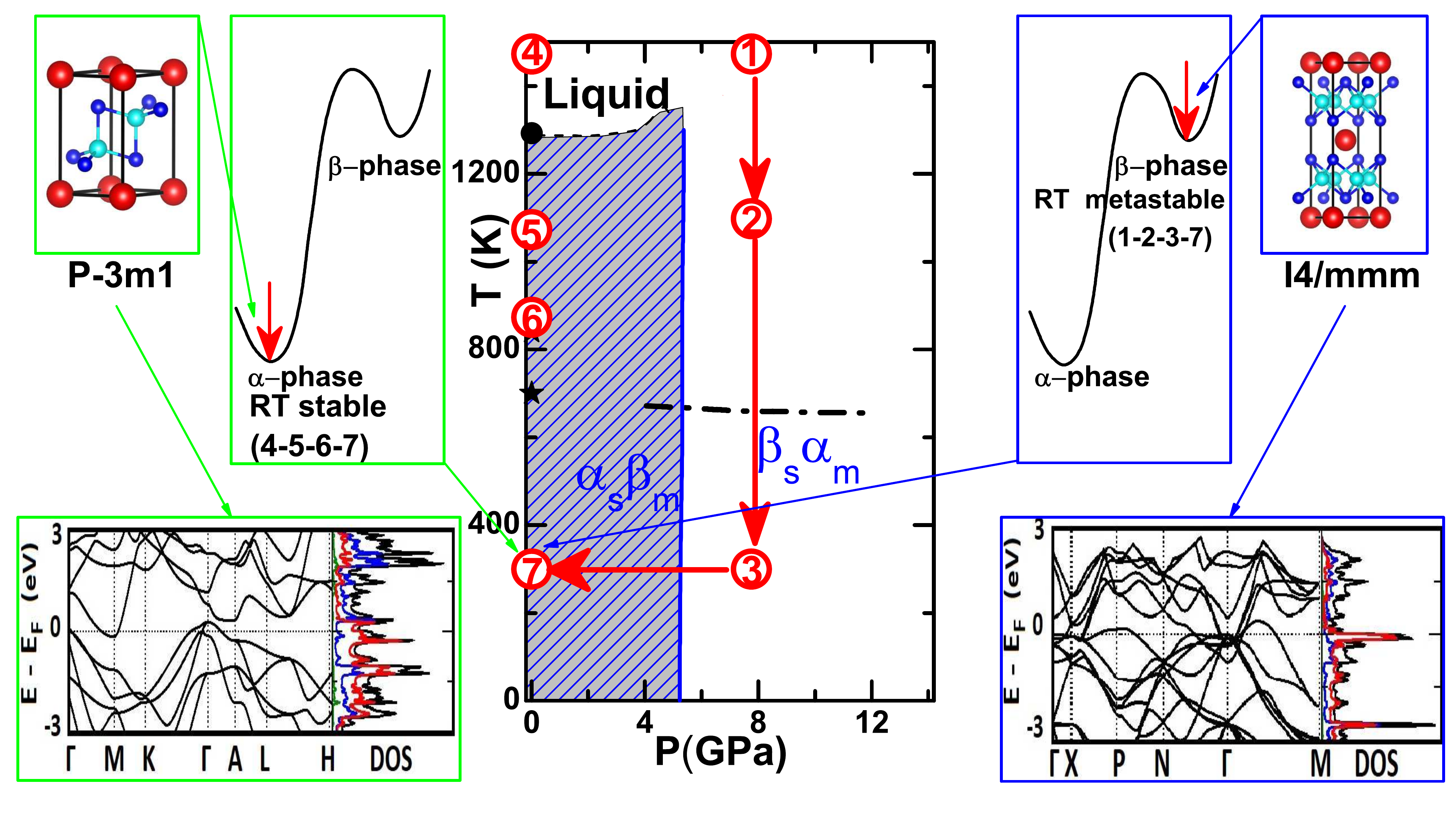}
\end{center}
\label{For Table of Contents Only}%
\end{tocentry}

\twocolumn[
\begin{@twocolumnfalse}
\begin{abstract}
SrAl$_2$Si$_2$ crystallizes into either a semimetallic, CaAl$_2$Si$_2$-type, $\alpha$ phase or a superconducting, BaZn$_2$P$_2$-type, $\beta$ phase.  We explore possible 
$ \alpha \xrightarrow {P_c,T_c} \beta$ 
transformations by employing  pressure- and temperature-dependent free-energy calculations, vibrational  spectra calculations, and room-temperature synchrotron X-ray powder diffraction (XRPD) measurements up to 14 GPa using diamond anvil cell. Our theoretical and empirical analyses together with all baric and thermal reported events on both phases allow us to construct a preliminary $P\text{\textendash} T$ diagram of transformations. 
Our calculations show a relatively low critical pressure for the $\alpha$ to $\beta$ transition (4.9 GPa at 0 K, 5.0 GPa at 300 K and 5.3 GPa at 900 K); nevertheless, our nonequilibrium analysis indicates that the low-pressure-low-temperature $\alpha$ phase is separated from metastable $\beta$ phase by a relatively high activation barrier.
This analysis is supported by our XRPD data at ambient temperature and $P \le 14$~GPa which shows an absence of $\beta$ phase even after a compression involving three times the critical pressure.  
Finally, we briefly consider the change in Fermi surface topology when atomic rearrangement takes place via either transformations among SrAl$_2$Si$_2$-dimorphs or total chemical substitution of Ca by Sr in isomorphous $\alpha$ CaAl$_2$Si$_2$; 
empirically, manifestation of such topology modification is evident when comparing the evolution of (magneto-)transport properties of members of SrAl$_2$Si$_2$-dimorphs and $\alpha$ isomorphs.
\end{abstract}
\end{@twocolumnfalse}
]

\section{Introduction \label{Sec.Introduction}}
Generally, a phase diagram of a polymorph exhibits a cascade of transformations among distinct phases \cite{Blank13-Phase-Transformation-Book,Brazhkin06-Metastability-Transitions-PhaseDiagram,Parija18-Metastable-Phase-Acessability-Utility,Recio15-High-Press-Sciences}. Based on consideration of energy landscape, some metastable phases are separated from the stable ones by a steep energy barrier such that a transformation among them can only be effected by energy-yielding control parameters such as pressure and temperature. On the other hand, consideration of kinetics enables one to identify non-equilibrium features such as (ir)reversibility of the transformation and (meta)stability of the induced phases. Various equilibrium and non-equilibrium techniques with adequate frequency and spacial resolution can be used to probe both the transformations and the induced phases. The obtained results are often compiled in, e.g. $P\text{\textendash} T$  diagram of transformations; usually these diagrams include adequate specification of characteristics such as (meta)stability, (ir)reversibility, and thermodynamic parameters (e.g. order and rate of transformation as well as the involved energetics).  \\
As an illustration, let us recall the transformations and phases encountered in the $P\text{\textendash} T$ phase diagram of carbon wherein the two allotropes (namely metastable diamond and stable graphite) are separated by a very high activation barrier.\cite{Bundy1989-T-P-Phase-Diagram-Carbon,Angus88-Diamond-Metastability} 
In contrast to these extreme conditions, there are various other transformations and induced phases which can be carried out under modest laboratory conditions.  Below, on employing experimental techniques (pressure-dependent structural analysis) and theoretical (equilibrium  free-energy  and  vibrational  spectra)  analysis, we are able to construct and analyze a $P\text{\textendash} T$ diagram of transformation of another example of a polymorphic system, namely the semimetallic layered \ce{SrAl2Si2}. As will be shown, much insight is gained on analyzing the (ir)reversibility of its structural transformations and the (meta)stability of its induced crystalline phases.

\begin{figure}[htbp] 
\centering
\includegraphics[width=1.0\columnwidth,keepaspectratio,trim={0.2cm 5cm 0.2cm 0.15cm}, clip]{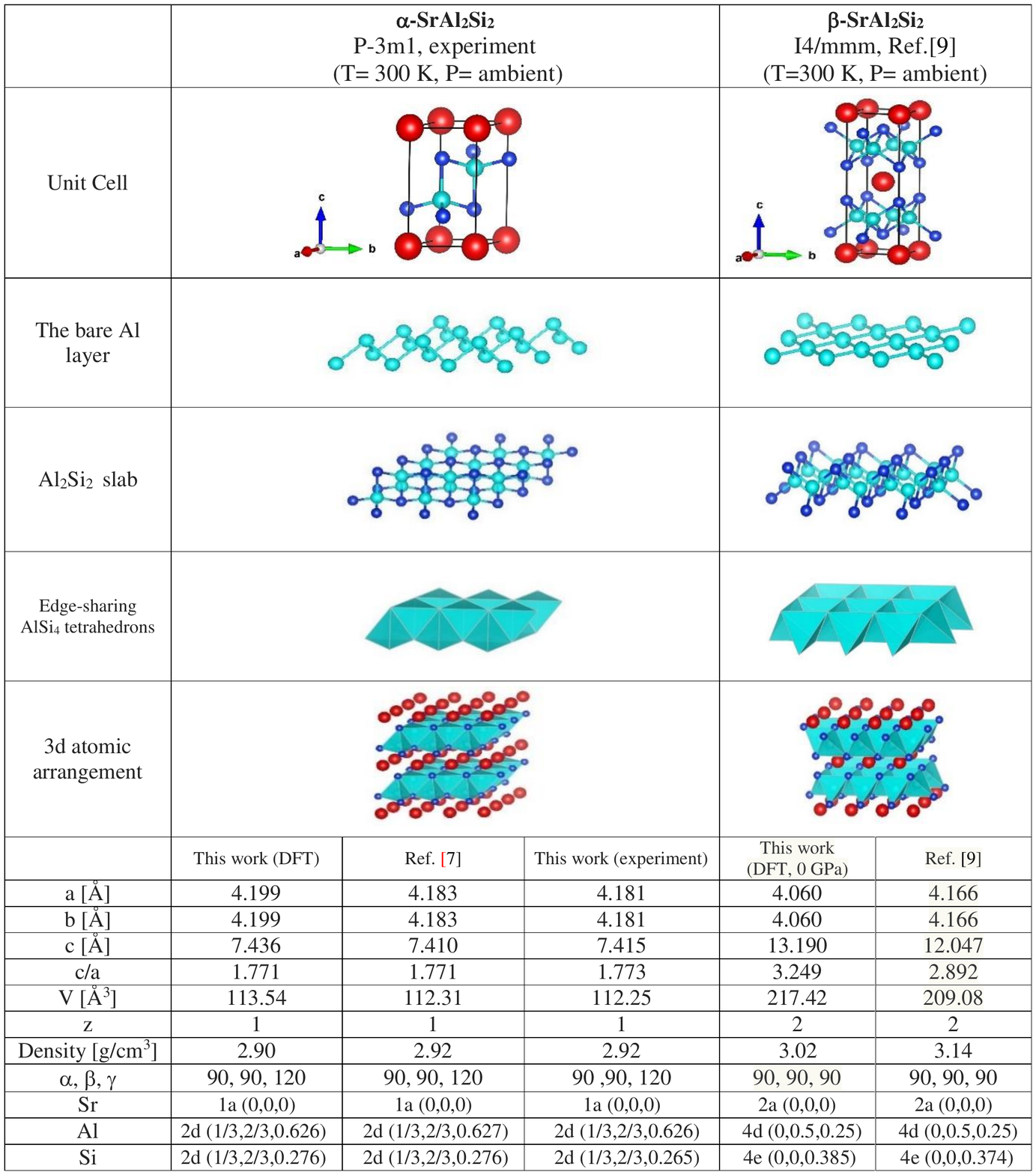}%
\caption{Comparison of the structural character of the $\alpha$  and $\beta$ phases of SrAl$_2$Si$_2$.  
For $\alpha$ SrAl$_2$Si$_2$, the structure parameters were taken from Ref.~7
, our experiments, and our optimized DFT-GGA analysis while the figures were adapted from 
Ref.~10
[Copyright (1988), with permission from Elsevier].
For $\beta$ SrAl$_2$Si$_2$, the structure parameters were taken from
Ref.~9
and our optimized DFT analysis while the figures were adapted from 
Ref.~7
[Copyright (2009), with permission from Elsevier] and Ref.~9
[Copyright (2017) American Chemical Society].  
For both phases, the GGA calculations gave, as expected, weakly enlarged volumes as well as weakly softer bulk moduli (Table~\ref{Table2-Bulk-Pc-Expt}) than the experimental values.\cite{Narasimhan02-LDA-GGA}
Additionally, DFT calculations at P=5 GPa indicate $a$=4.110~\AA~ and $c$= 7.210~\AA~ for $\alpha$ SrAl$_2$Si$_2$, while $a$=4.102~\AA~ and $c$=11.177~\AA~ for $\beta$ SrAl$_2$Si$_2$. 
Bonding properties of each constituent ion were amply discussed in 
Refs.7,9 
. Symbols are as follows: Sr $\equiv$ large-size red circle, Al $\equiv$ intermediate-size cyan circle, 
Si $\equiv$ small-size blue circle. All drawings of crystal structures were produced via VESTA program \cite{Momma-db5098}.
} 
\label{Fig1-Sr-Table-Trig-Tetra}%
\end{figure}

SrAl$_2$Si$_2$ crystallizes into two layered phases (see Fig.~\ref{Fig1-Sr-Table-Trig-Tetra}). The first, so-called $\alpha$ phase, exhibits a stable trigonal ($P\bar{3}m1$) structure wherein slabs of three-edges-sharing tetrahedraly-coordinated AlSi$_4$ are intercalated, along the $c$-axis, with Sr planes \cite{kauzlarich09-SrAl2SI2-structure-Thermoelectric,Lue11-Electronic-structure-(Sr-Y)Al2Si2, Zevalkink17-SrAl4-xSix-Making-Breaking-Bonds}. The related electronic structure exhibits a series of hole and electron pockets contributing to a total density of states \big(DOS at E$_F$, $N_t(E_F)$\big) which is relatively lower than the one observed in isomorphous $\alpha$ \ce{CaAl2Si2} \cite{Lue11-Electronic-structure-(Sr-Y)Al2Si2, Zevalkink17-SrAl4-xSix-Making-Breaking-Bonds}. The observed electrical transport properties were reported to be non-superconducting semimetallic with dominant contribution from electron charge carriers \cite{kauzlarich09-SrAl2SI2-structure-Thermoelectric,Lue11-Electronic-structure-(Sr-Y)Al2Si2, Zevalkink17-SrAl4-xSix-Making-Breaking-Bonds}. 

The second, so called $\beta$ phase, is metastable at low pressures and exhibits BaZn$_2$P$_2$-type ($I4/mmm$) layered structure wherein slabs of four-edges-sharing AlSi$_4$ are intercalated, along the $c$-axis, with Sr planes \cite{Zheng88-Studies-on-ThCr2Si2-CaAl2Si2} (see also Fig.~\ref{Fig1-Sr-Table-Trig-Tetra}). Its electronic structure is metallic, with predominantly electronic charge carriers, and exhibits a relatively higher $N_t(E_F)$; this is assumed to be the driving ingredient behind the surge of superconductivity below $T_c \approx$ 2.6~K \cite{Zevalkink17-SrAl4-xSix-Making-Breaking-Bonds}. 

The $\beta$ phase can be stabilized by quenching of the high-pressure and high-temperature (HPHT) phase into ambient conditions \cite{Zevalkink17-SrAl4-xSix-Making-Breaking-Bonds}. However, this \textcolor{red}{$\beta$} phase is metastable as it irreversibly transforms \textcolor{red}{back} into the $\alpha$ phase upon heating under ambient pressure \cite{Zevalkink17-SrAl4-xSix-Making-Breaking-Bonds}.

\begin{figure}[htbp] 
\centering
\noindent\includegraphics[width=0.8\columnwidth,trim={0cm 0cm 0cm 0cm}, clip]{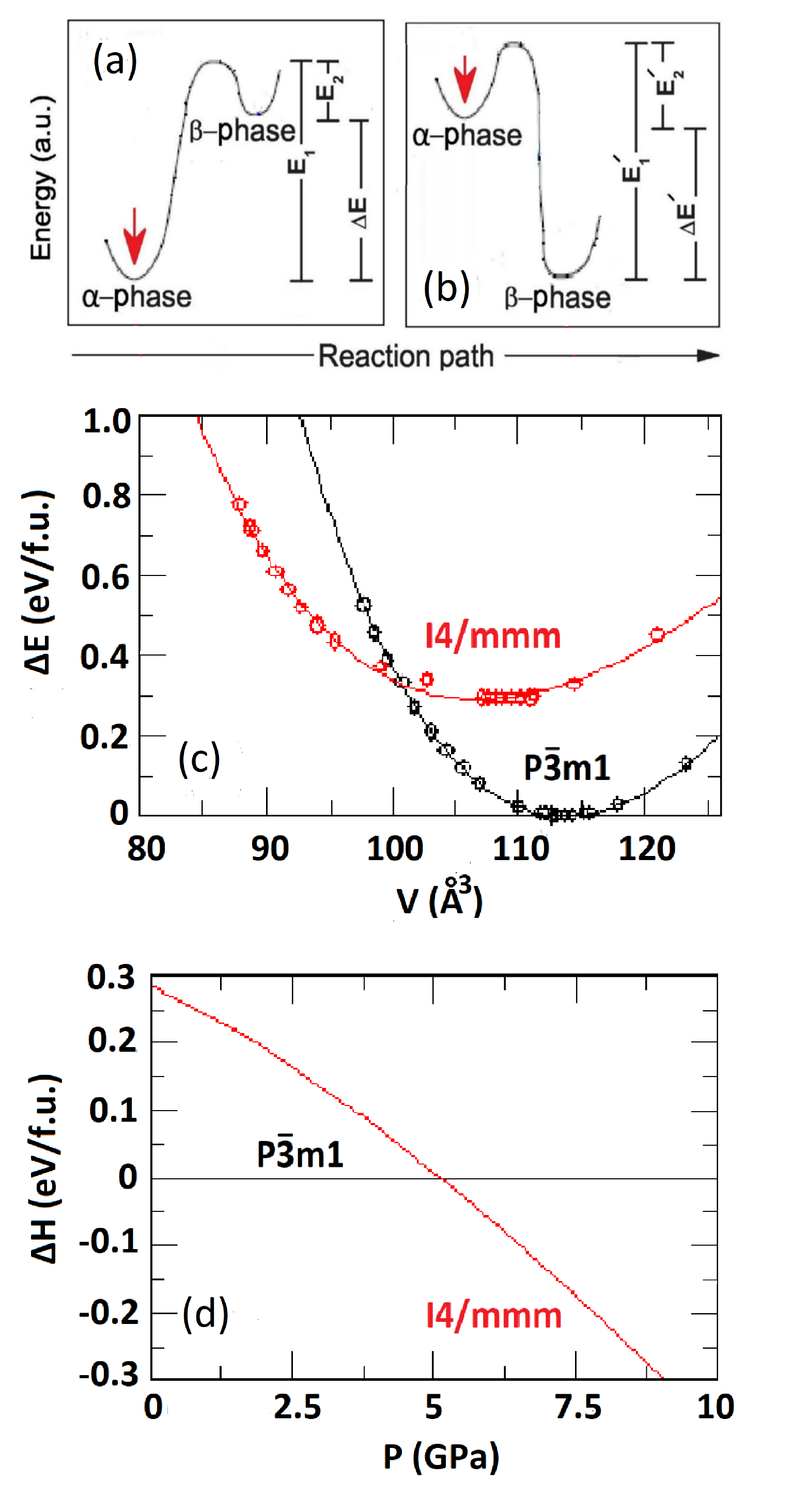}%
\caption{Free energies of SrAl$_2$Si$_2$ dimorphs. \textit{(a)} Energy as a function of reaction path: 
$\alpha$ phase is at a global minimum (similar to the initial LPLT phase in Fig.~\ref{Fig9-Sr-P-T-Phase-Diagram-Proposed}).
\textit{(b)} $\beta$ phase is at a global minimum while $\alpha$ phase is metastable (similar to the measured HPLT phase in Fig.~\ref{Fig9-Sr-P-T-Phase-Diagram-Proposed}, $P>P_c^{cal}$ and $k_B T<E^\prime_2$).
\textit{(c)} Relative internal energy \textit{vs} volume, $\Delta E(V)$, of SrAl$_{2}$Si$_{2}$. The solid lines are  calculated from Murnaghan equation (Eq.\ref{Eq-Murnaghan-E-vs-V}) for the two structural phases.
\textit{(d)} Relative enthalpy \textit{vs} pressure, $\Delta H(P)$, curves with respect to the enthalpy of the most stable structure at zero pressure. The solid lines are calculated via Eq.\ref{Eq-Enthalpy-vs-P}. 
} \label{Fig2-Srl2Si2-Barrier-Phases-E-V-H-P-curves}%
\end{figure}

The atomic arrangement in these dimorphous forms are not similar and any $\alpha \rightarrow \beta$ transition would take place only after overcoming a high energy barrier (see Fig.~\ref{Fig2-Srl2Si2-Barrier-Phases-E-V-H-P-curves}(a,b)); only then Sr coordination changes from being at the center of a trigonal anti-prismatic cage into being at the center of square prismatic one; Al within the Al$_2$Si$_2$ slab changes from an AB-close-packed double-sheet into a planar square net; and the intralayer coordination of Si transforms from an umbrella-type into a square-pyramid one. 

In this work, we show that all of the above empirical results (namely, HPHT synthesis routes, \cite{Zevalkink17-SrAl4-xSix-Making-Breaking-Bonds} together with the structural and thermal events\cite{Lue11-Electronic-structure-(Sr-Y)Al2Si2,kauzlarich09-SrAl2SI2-structure-Thermoelectric, Zevalkink17-SrAl4-xSix-Making-Breaking-Bonds}) can be elegantly compiled into a single $P\text{\textendash} T$ diagram of transformations.  
Moreover, on the top of this empirical diagram, we superimpose another diagram composed out of the results obtained from room-temperature pressure-dependent synchrotron X-ray powder diffraction, XRPD, analysis and the pressure- and temperature-dependent free-energy calculations (in particular the prediction of reversible transitions and low-energy structural phases).
The final construction is an equilibrium and nonequilibrium  $P\text{\textendash} T$ diagram of transformations that highlights the character of the structural phases and the transformations (in particular the $P_c(T_c)$ boundaries that delineate the $\alpha \xrightarrow[]{P_c,T_c} \beta$ transitions).

%
It is expected that the inter-transformation among SrAl$_2$Si$_2$-dimorphs or complete chemical substitution within the $\alpha$ $A$Al$_2$Si$_2$ ($A$=Ca, Sr) isomorphs involves extensive atomic rearrangements or lattice parameters variation. This, in turn, is expected to have a direct impact on the related electronic structures. We are particularly interested in the evolution of the topology of the Fermi surfaces across the SrAl$_2$Si$_2$ dimorphs and $\alpha$ $A$Al$_2$Si$_2$ isomorphs as well as in how these modifications
would be reflected in the (magneto-) transport properties. As an illustration, let us recall two features which are strikingly different among SrAl$_2$Si$_2$ dimorphs \cite{kauzlarich09-SrAl2SI2-structure-Thermoelectric,Lue11-Electronic-structure-(Sr-Y)Al2Si2, Zevalkink17-SrAl4-xSix-Making-Breaking-Bonds}: the superconductivity (absent in $\alpha$ phase but emerging in $\beta$ phase) and the sign of charge carriers (mixed in $\alpha$ phase while predominantly negative in $\beta$ phase). 
As a further illustration, let us compare the electronic properties of the two
$\alpha$ $A$Al$_2$Si$_2$ isomorphs: as that the evolution of transport properties of both compounds reflects contributions from hole and electron pockets \cite{kauzlarich09-SrAl2SI2-structure-Thermoelectric,Lue11-Electronic-structure-(Sr-Y)Al2Si2, Zevalkink17-SrAl4-xSix-Making-Breaking-Bonds,Costa18-CaAl2Si2-MagRes}, it is interesting to investigate whether the reported robust linear magnetoresistivity (LMR) of $\alpha$ CaAl$_2$Si$_2$ \cite{Costa18-CaAl2Si2-MagRes} is also manifested in isomorphous $\alpha$ SrAl$_2$Si$_2$. As far as the superconducting $\beta$ SrAl$_2$Si$_2$ is concerned, this LMR feature is totally suppressed within the studied experimental conditions  \cite{Zevalkink17-SrAl4-xSix-Making-Breaking-Bonds}.

This article is formatted as follows. In $\S$\ref{Sec.Methods-Theory-Experiment}, we present the employed theoretical and experimental methods. 
In $\S$\ref{SubSec-Theoretical-Results}, we present zero-temperature and finite-temperature calculations.
We show, in $\S$\ref{Experimental-Results}, our room-temperature pressure-dependent diffractograms and ambient-pressure temperature-dependent magnetoresistivity of $\alpha$ SrAl$_2$Si$_2$. 
We emphasize, in particular, that the experimental XRPD data do not evidence any $\alpha \rightarrow \beta$ transition up to 14 GPa, in contrast to the computational results; reconciliation of this discrepancy between prediction and experiment will be discussed in terms of nonequilibrium $P\text{\textendash} T$ diagram which is constructed in $\S$\ref{SubSubSec.Phase-Diagram}. Discussion and conclusion are given in $\S$\ref{Sec.Discussion-Conclusion}.   

\section{Methods \label{Sec.Methods-Theory-Experiment}}
\subsection{Theoretical Techniques \label{SubSec.Theory-procedure}} 
Using the Quantum Espresso package \cite{giannozzi09-espresso}, we initially performed full geometry optimizations 
for different target pressures ($P\leq 10$~GPa) and zero temperature. 
Our calculations are based on plane-wave basis sets and pseudopotentials. Ultra-soft Vanderbilt pseudopotentials \cite{Vanderbilt90-pseudopotentials}, plane wave cutoff energy of 50 Ry and GGA-PBE exchange correlation functional \cite{perdew96-GGA} were used. We considered 10, 3 and 4 valence-electrons pseudopotentials for Sr, Al, and Si, respectively. 

The electronic band structure calculations were performed after geometry optimizations. The energy convergence criterion for the self-consistency loop was set to $10^{-7}$  eV and the lattice parameters and atomic positions were optimized until the forces on all atoms were smaller than $10^{-3}$ eV/Å. The electronic k-point mesh chosen was 10 × 10 × 10.

The stability of the SrAl$_2$Si$_2$ dimorphs was verified by comparing the evaluated internal energy versus volume, $E(V)$, and enthalpy versus pressure, $\Delta H (P)$, of the most-probable structural candidates \cite{Note-Four-Possible-phases}.
All $E(V)$ curves are fitted to the Murnaghan equation-of-state \cite{Murnaghan37-finite-deformations}: 
\begin{equation}
 \begin{aligned}
  E(V) = {} & B_{0}V_0 \Big[\frac{1}{B_0'(B_0'-1)} \Big(\frac{V_0}{V} \Big)^{B_0'-1} \\
            &  + \frac{V}{B_0'V_0} - \frac{1}{B_0'-1} \Big]\
  \end{aligned}
\label{Eq-Murnaghan-E-vs-V}
\end{equation} %
In Eq.~\ref{Eq-Murnaghan-E-vs-V}, $E$ is internal energy, $V$ is volume, B$_{0}$ and B$^\prime_{0}$ are the bulk modulus and its derivative with respect to pressure respectively; the subscript 0 denotes the zero-pressure values. 

Following the procedures implemented in the Quantum Espresso code \cite{Baroni_2010}, we extracted elastic constants C$_{ij}$, after structure optimization, from the second derivative of the total energy with respect to the strain along two crystallographic directions:
\begin{equation}
C_{ij} = \frac{1}{V}\frac{\partial^2 E(V)}{\partial\varepsilon_i\partial\varepsilon_j}.
 \label{elastic-constant}
\end{equation}
For hexagonal and tetragonal crystals, there are six independent elastic constants based on Eq.~\ref{elastic-constant}: (C$_{11}$, C$_{12}$, C$_{13}$, C$_{33}$, C$_{44}$, C$_{66}$) and (C$_{11}$, C$_{12}$, C$_{13}$, C$_{14}$, C$_{33}$, C$_{44}$), respectively. 
We examined the pressure-dependence of these elastic constants (as well as the bulk moduli) at $P=$0 and 5~GPa. 

We also derived the pressure-dependent enthalpy, $H(P)$,:
\begin{equation}
 H(P)= \frac{B_0V_0}{B_0'-1}\Big[ \Big( \frac{B_0'P}{B_0}+1\Big)^{1-\frac{1}{B_0'}}-1\Big],
\label{Eq-Enthalpy-vs-P}
\end{equation}
and the $V(P)$ relation
\begin{equation}
V(P)=V_{0}\left[1+P\left({\frac  {B^\prime_{0}}{B_{0}}}\right)\right]^{-1/B^\prime_0},
\label{Eq-V-vs-P-Murnaghan}
\end{equation}
which is used in $\S$~\ref{subSec.Sr-Results} for comparing theory and experiment. 

Finally, we performed DFT-based calculations of vibrational (phonon) spectra to compute vibrational contributions to the free energy of the SrAl$_2$Si$_2$ dimorphs, at various pressures. The quasi-harmonic approximation was used within the Quantum Espresso code  \cite{Baroni_2010}.
From these calculations, we extract the zero-point energy, entropy, and Gibbs free energy using standard expressions \cite{Baroni_2011}.
The free-energy analysis allows us to calculate the corresponding equilibrium $P\text{\textendash} T$ phase diagram. 
\begin{table*}
\scriptsize
\caption{Elastic constants (in [GPa]) and average Debye sound velocities of SrAl$_2$Si$_2$ dimorphs, under two representative pressures, as calculated via the DFT-GGA approximation. One verifies that the necessary and sufficient conditions for elastic stability for the two dimorphs, Eq.~(\ref{Eq-Elastic-Stability}), are satisfied.}
\begin{tabular}[c]{cccccccccccc}
\hline\hline
($GPa$) & & B$_{0}$ [GPa] & $B^\prime _{0}$& v(m/s)&C$_{11}$ & C$_{12}$ & C$_{13}$ & C$_{33}$ & C$_{14}$ & C$_{44}$ & C$_{66}$ \\ %
\hline
 $P\bar{3}m1$, 0~GPa & & 63.2& 3.9 &4538.76&141.7 & 26.5 & 26.4 & 129.9 & 0 & 40.3 & - \\ %
 $P\bar{3}m1$, 5~GPa & & 83.7 &3.8 &4889.07&177.1 & 40.0 & 39.0 & 163.9 & 0 & 52.9 & - \\ %
 $I4/mmm$,     0~GPa & & 62.0 &4.4 &3324.95&115.7 & 39.6 & 42.3 &  76.1 & - & 24.7 & 27.3 \\ %
 $I4/mmm$,    5~GPa & &84.8 & 4.2&3777.31 &132.5 & 73.0 & 63.0 & 111.6 & - & 46.9 & 59.2\\ %
\cline{1-12}
\end{tabular}
\label{Table1-ElasticConstant}%
\end{table*} 

\begin{table*}
\scriptsize
\caption{Experimental linear compressibilities ($\frac{\delta(a/a_{o})}{\delta P}$,  $\frac{\delta(c/c_{o})}{\delta P}$) and bulk moduli and their derivatives (B$_{0}$ in [GPa] and  B$^\prime_{0}$), at T=300~K, of $\alpha$ SrAl$_2$Si$_2$. $\frac{\delta(a/a_{o})}{\delta P}$ and $\frac{\delta(c/c_{o})}{\delta P}$ (in TPa$^{-1}$) were obtained from a linear fit, close to P=0~GPa, of the data depicted in Figs.~\ref{Fig7-Sr-Param-2nd-run}(a,b). All quantities are compared with the corresponding theoretical values obtained at P=0 GPa (See also Table~\ref{Table1-ElasticConstant}). The strong correlation between B$_{0}$ and B$^\prime_{0}$ parameters is depicted in the form of a confidence ellipse in Fig.~\ref{Fig7-Sr-Param-2nd-run}(e). As the structure is more compressible along the $c$-axis, then the corresponding  $\lvert \frac{\delta(c/c_{o})}{\delta P} \rvert$ is higher than that along other axes [see text and Eqs.~(\ref{Eq-Compressibilities-along-a},\ref{Eq-Compressibilities-along-c})]. Also shown are the predicted structural transformation and calculated critical pressure, $P_c^{cal}$ in [GPa].}  
\begin{tabular}[c]{ccccccccccc}
\hline\hline
 & \multicolumn{2}{c} {$\frac{\delta(a/a_0)}{\delta P}$}  & \multicolumn{2}{c} {$\frac{\delta(c/c_0)}{\delta P}$}   & \multicolumn{2}{c}{B$_{0}$ } &  \multicolumn{2}{c} {$B^\prime _{0}$} & $P_c^{cal}$ & $\alpha \rightarrow \beta$ Transition\\
- & Exp. &Theor. & Exp. &Theor. & Exp. &Theor. & Exp. &Theor. & & \\ %
\hline 
Sr  & -2.9(1)& -5.0 & -4.1(1) & -5.6 & 64(9)& 63.2 &5(1) & 3.9 & 5.0&$P\bar{3}m1\rightarrow I4/mmm$\\ %
\hline 
\end{tabular}
\\
\label{Table2-Bulk-Pc-Expt}%
\end{table*} 

\subsection{Experimental Techniques \label{SubSec.Experiment-procedure}}
Sample preparation and experimental setups are the same as the ones described in Ref.~\onlinecite{Strikos20-CaAl2Si2-Metastability}. Polycrystalline samples were synthesized via standard Argon arc-melt procedure using high purity elements. 

Diffraction patterns were collected \textit{in situ} under high pressures exerted through a diamond anvil cell, at Xpress beamline - Elettra Sincrotrone Trieste, Italy \cite{Lotti20-Single-crystal-diffraction-Elettra} - using a monochromatic beam of $\lambda=0.4957$~\AA. More details on the procedures adopted for high pressure diffraction measurements is described in Ref.~\onlinecite{Strikos20-CaAl2Si2-Metastability}.

Structural analyses were carried out using the Rietveld refinement as implemented in the Fullprof package \cite{Rodriguez01-Program-FULLPROF}. During the refinement, the Wyckoff positions and site occupancy were fixed or varied depending on the conditions given in Space Group Tables. In particular, only the following parameters were allowed to vary: z of Al and Si in $P\bar{3}m1$ and z of Si in $I4/mmm$. The sites preference and atomic displacement parameters were taken from the respective references quoted in Fig.~\ref{Fig1-Sr-Table-Trig-Tetra}.

Standard magnetization measurements were performed on a Vibrating Sample Magnetometer (VSM) while the electrical (magneto-)resistivity measurements on a standard four-probe dc-driven setup. Both were implemented in the Physical Property Measurement System (PPMS) of Quantum Design environment with a temperature range of 2 to 300~K and a magnetic field up to 90~kOe.

\section{Results and Analysis \label{Sec.Results}}
\subsection{Theoretical Results\label{SubSec-Theoretical-Results}}

\begin{figure}[htbp]
\centering
\includegraphics[width=0.7\columnwidth,trim={0.0cm 0.0cm 0cm 0cm}, clip]{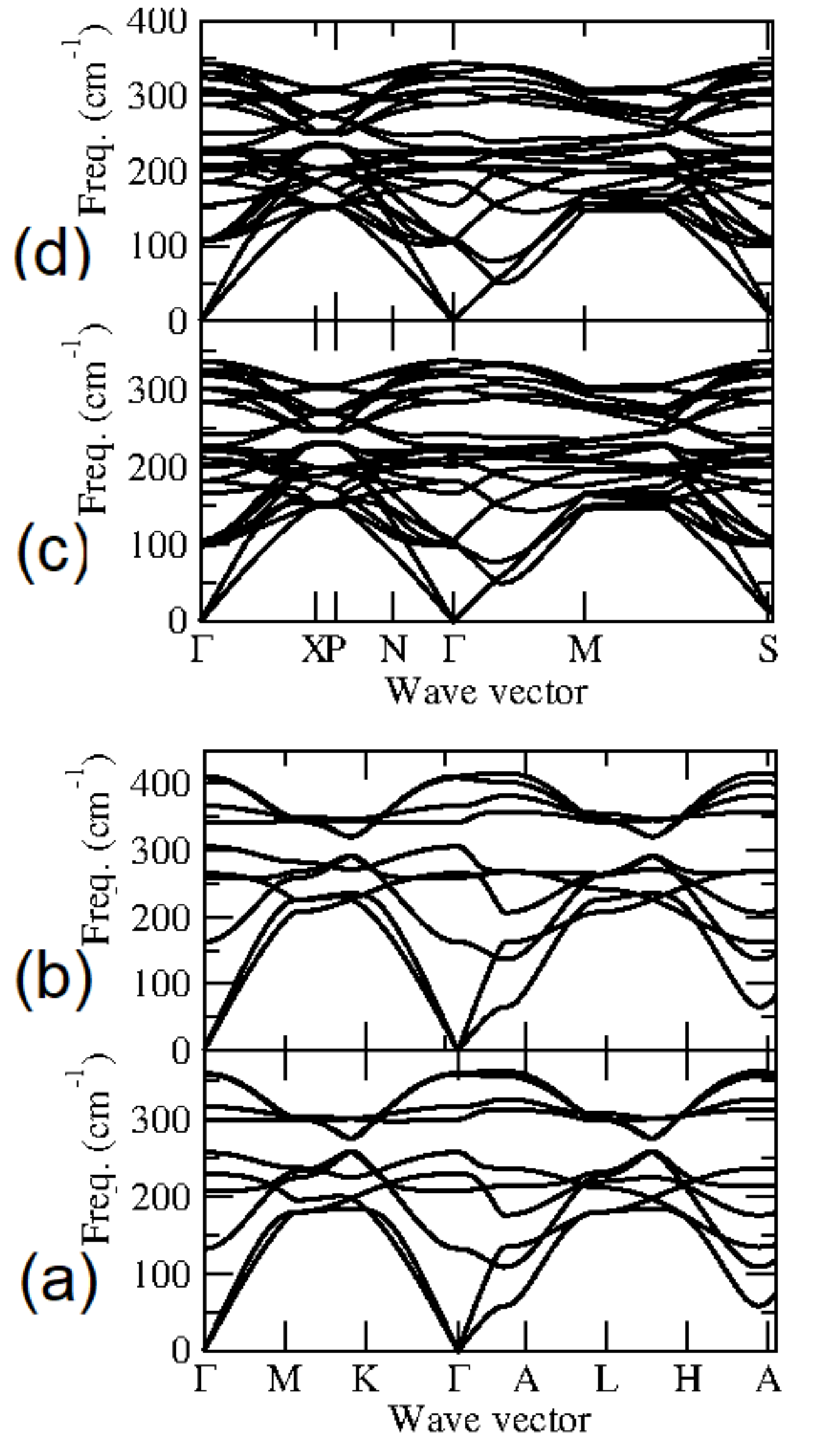}%
\caption{ 
Vibrational dispersion curves of both $\alpha$  and $\beta$ phases of SrAl$_2$Si$_2$ for two representative pressures: \textit{(a,~b)} $\alpha$ phase at $P=$0~GPa and $P=$8~GPa respectively. Similarly, \textit{(c,~d)} represent the corresponding curves of the $\beta$ phase for both $P=$0~GPa and $P=$8~GPa, respectively.
}
\label{Fig3-Phonon-Spectra}%
\end{figure} 
%
\begin{figure}[htbp] 
\centering
\includegraphics[width=1.0\columnwidth,trim={0.0cm 0.00cm 0cm 0cm}, clip]{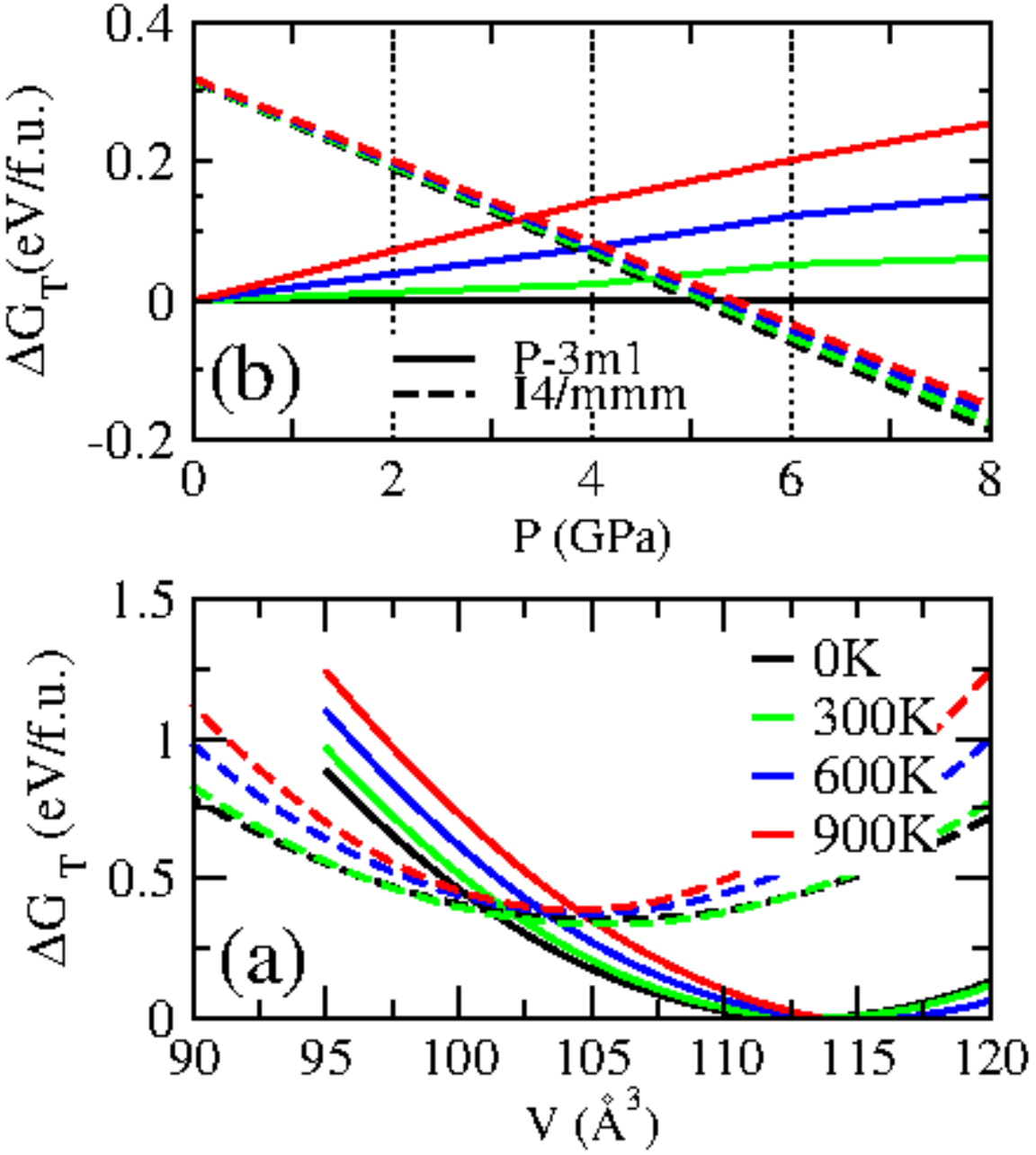}%
\caption{ 
Evolution of Gibbs energies and the phase diagram  of SrAl$_2$Si$_2$. (a) Calculated isothermal Gibbs energy $versus$ volume, $\Delta G_T (V)$, of $\alpha$ (solid lines) and $\beta$ (dashed lines) phases. Similarly, (b) shows $\Delta G_T (P)$ of $\alpha$ (solid lines) and $\beta$ (dashed lines) phases, using the same color code as in panel (a). Calculations were undertaken for $T=$ 0, 300, 600 and  900 K. Energy reference was chosen to be that of $\alpha$ phase at zero-temperature. Based on the analysis of the curves in panel (b), one obtains the temperature-dependent critical pressure of $\alpha \xrightarrow {P_c(T_c)} \beta$ transition and therefrom constructs the $P\text{\textendash} T$ phase diagram shown in Fig.~\ref{Fig9-Sr-P-T-Phase-Diagram-Proposed}. 
}
\label{Fig4-GvsP-Phase-Diagram}%
\end{figure} 
%
\subsubsection{$T = 0$~K and $P\leq 10$~GPa calculations \label{SuSubSec-T=0-Calculation}}
Figures~\ref{Fig2-Srl2Si2-Barrier-Phases-E-V-H-P-curves} (c,~d) show the internal  energy  versus  volume ($E(V)$ according to Eq.~\ref{Eq-Murnaghan-E-vs-V}) and enthalpy versus pressure ($\Delta H (P)$ according to Eq.~\ref{Eq-Enthalpy-vs-P}) for both the tetragonal $I4/mmm$ \cite{Yamanaka04-BaAl2Si2} and the trigonal $P\bar{3}m1$ \cite{Gladyshevsky67-Cryst-Str,Note-Four-Possible-phases} phases.
In agreement with experiments, these figures indicate that, at ambient-pressure, SrAl$_2$Si$_2$ crystalizes in the hexagonal P$\overline{3}$m1 phase. On the other hand, the I4/mmm phase is the most stable one at sufficiently high pressures.
In addition, on comparing the relative enthalpies of the studied phases, we are able to identify  the critical pressure at which $\alpha \xrightarrow {P_c^{cal}} \beta$ transition occurs.
The calculated $P_c^{cal} \approx 5.0$ GPa falls, as discussed in $\S$~\ref{SubSubSec.Phase-Diagram}, in the range of pressures employed during the \textit{hot-compression cycle} to promote the $\alpha \xrightarrow[]{HPHT} \beta$ transformation (4 $< P \le$ 9.5~GPa and 400 $\leq T\leq $ 1250 °C,~\cite{Zevalkink17-SrAl4-xSix-Making-Breaking-Bonds,Zevalkink20-SrAl2Si2-PrivateCom}).

Table~\ref{Table1-ElasticConstant} summarizes the calculated bulk modulus, its derivative, and the elastic constants at selected pressures. 
Two points are worth mentioning: First, we verified that the mechanical stability criteria~\cite{mouhat14-elastic-stability-conditions}
\begin{equation}
\begin{split}
C_{11} >\lvert C_{12} \rvert, \, \, \,  C_{44}> 0, \\ 
2C_{13}^{2}< C_{33}(C_{11} + C_{12}), \, \, \,  C_{66}> 0
\label{Eq-Elastic-Stability}
\end{split}
\end{equation} 
are satisfied for both phases.
Second, the observed pressure-induced increase in bulk modulus is indicative of an increased structural rigidity. But, as SrAl$_2$Si$_2$ is a layered structure, wherein the covalently-bond Al$_2$Si$_2$ slabs are stacked along the $c$-axis (Fig.~\ref{Fig1-Sr-Table-Trig-Tetra}), the baric evolution of their lattice parameters is expected to be anisotropic. Indeed, the calculated elastic constants indicate that C$_{11}>$ C$_{33}$, i.e. compression along [001] direction is easier than along [100].

It is instructive to convert these elastic constants into linear compressibilities, since the latter are a useful measure of the anisotropy in the strength of intermolecular interactions \cite{stevens05-elastic-constants}. Then, we are able to compare these theoretically-obtained compressibilities with the corresponding empirically-determined ones. For the hexagonal $\vec{a}$ and $\vec{c}$ directions, the following relations were employed \cite{nye85-physical-properties-crystals,kimizuka07-complete-elastic-const-quartz}: 
\begin{subequations}
\begin{flalign} \label{Eq-Compressibilities-along-a}
\frac{\delta(a/a_o)}{\delta P}  = -\frac{C_{33}-C_{13}}{C_{33}(C_{11}+C_{12})-2C_{13}^2} = -\beta_{\alpha} \\
\frac{\delta(c/c_o)}{\delta P}  = -\frac{C_{11}+C_{12}-2C_{13}}{C_{33}(C_{11}+C_{12})-2C_{13}^2} =  -\beta_{c}
\label{Eq-Compressibilities-along-c} 
\end{flalign}
\end{subequations}%
The converted compressibilities as well as the experimental ones are shown in Table~\ref{Table2-Bulk-Pc-Expt} and will be discussed in 
$\S$~\ref{subSec.Sr-Results}.

\subsubsection{$0 < T \leq 900$~K and $P\leq 8~\text{GPa}$  calculations \label{SuSubSec-T>0-Calculation}}
Representative vibrational dispersion curves, within $P\leq$8~GPa, for $\alpha$ phase are shown in Figs.\ref{Fig3-Phonon-Spectra}(a,~b) while that for $\beta$ phase is in Fig.\ref{Fig3-Phonon-Spectra}(c,~d). From the absence of imaginary frequencies in the Brillouin Zone, it is clear that both phases are dynamically stable within the studied pressure range. From the linear dispersion of the longitudinal acoustic modes, we calculate and show in Table~\ref{Table1-ElasticConstant} the average Debye sound velocities of the two dimorphs. As expected, the sound velocities are faster in the $\alpha$ phase because its bulk modulus (Table~\ref{Table1-ElasticConstant}) is higher and its density (Fig.~\ref{Fig1-Sr-Table-Trig-Tetra}) is lower.

Figure~\ref{Fig4-GvsP-Phase-Diagram}(a) shows the calculated free energies at various fixed temperatures, $ \Delta G_T \; vs\; V$, while Fig.~\ref{Fig4-GvsP-Phase-Diagram}(b) shows the corresponding $ \Delta G_T \; vs\; P$ curves. The latter ones allow us to obtain the temperature-dependent critical pressure of $\alpha \xrightarrow {P_c(T_c)} \beta$ transition  and from there we construct the predicted $P\text{\textendash} T$ phase diagram shown in Fig.~\ref{Fig9-Sr-P-T-Phase-Diagram-Proposed} (the thick blue solid line). Within this equilibrium analysis, the stable structure for $P < P_c(T_c)$ (blue hatched area) is the $\alpha$ phase whereas for $P> P_c(T_c)$ it is the $\beta$ phase. 
We observe a monotonically increasing temperature-dependence of the critical pressure. The consequence of this distinct curve (as well as its distinct $\frac{dP}{dT}$) will be discussed in $\S$ \ref{SubSubSec.Phase-Diagram}.
%

\begin{figure}[htbp] 
\centering
\noindent\includegraphics[scale=0.4]{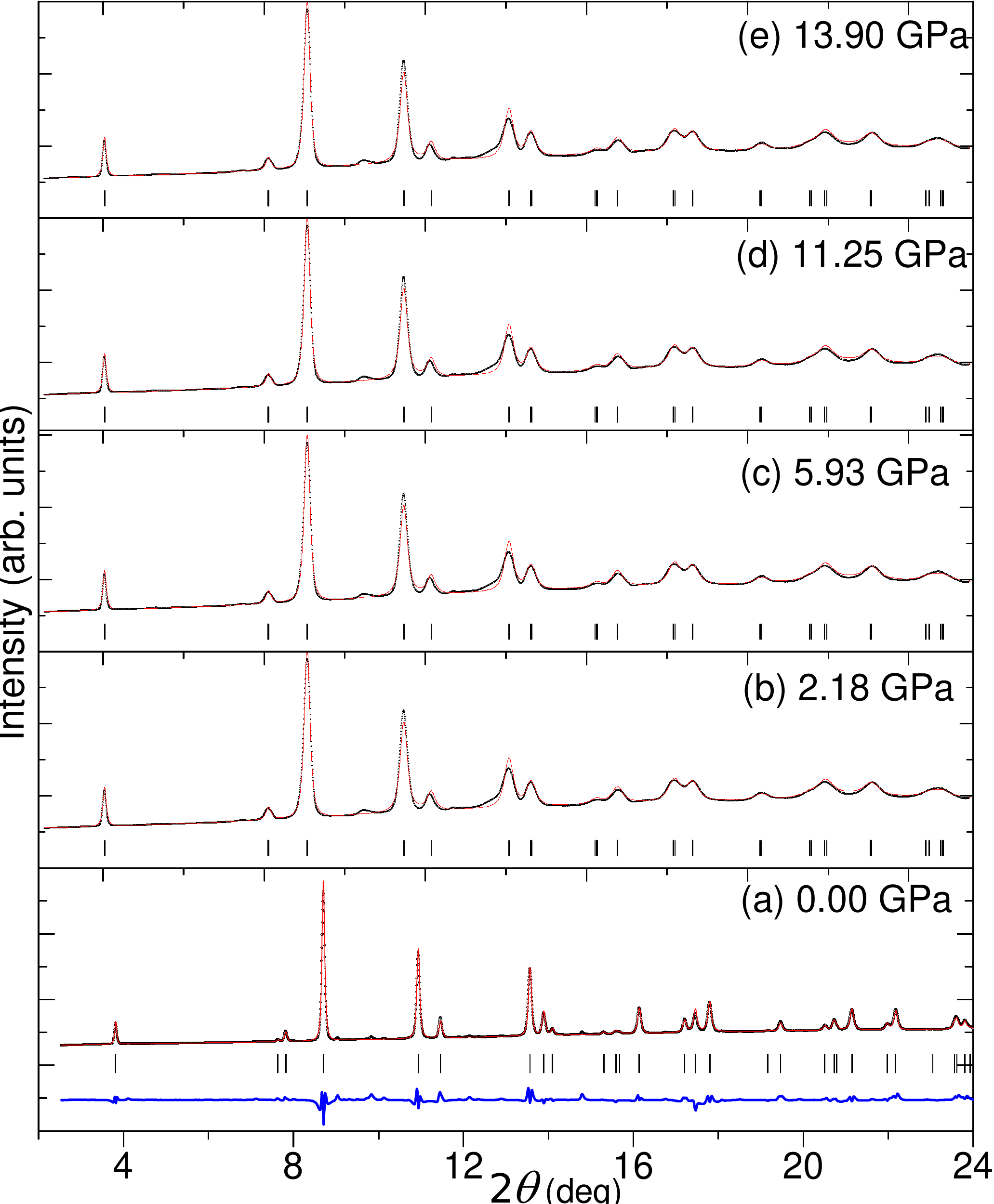}%
\caption{Selected pressure-dependent room-temperature diffractograms of $\alpha$ SrAl$_2$Si$_2$ for the following values: (a) ambient pressure, (b) 2.18 GPa, (c) 5.93 GPa, (d) 11.25 GPa, and (e) 13.90 GPa. \textit{Symbols:} the measured intensities as symbols, 
Bragg positions as short vertical bars, and calculated intensities as continuous red lines. The difference between the measured and calculated intensities (as a blue line) is shown only in (a). 
As evident, the Bragg peaks at ambient pressure are sharper than those at higher pressures: Fig.~S1 of the Supporting Information shows the baric evolution of both the full-width at half maximum and the area of (011)/(101) Bragg peak. 
Moreover, Table S1 shows the reliability factors $\chi^{2}$, $R_{p}$, $R_{wp}$ for the studied diffractograms.
} \label{Fig6-Sr-Diffractogram-2nd-run}
\end{figure}

\begin{figure}[htbp]
\centering
\noindent\includegraphics[scale=0.3]{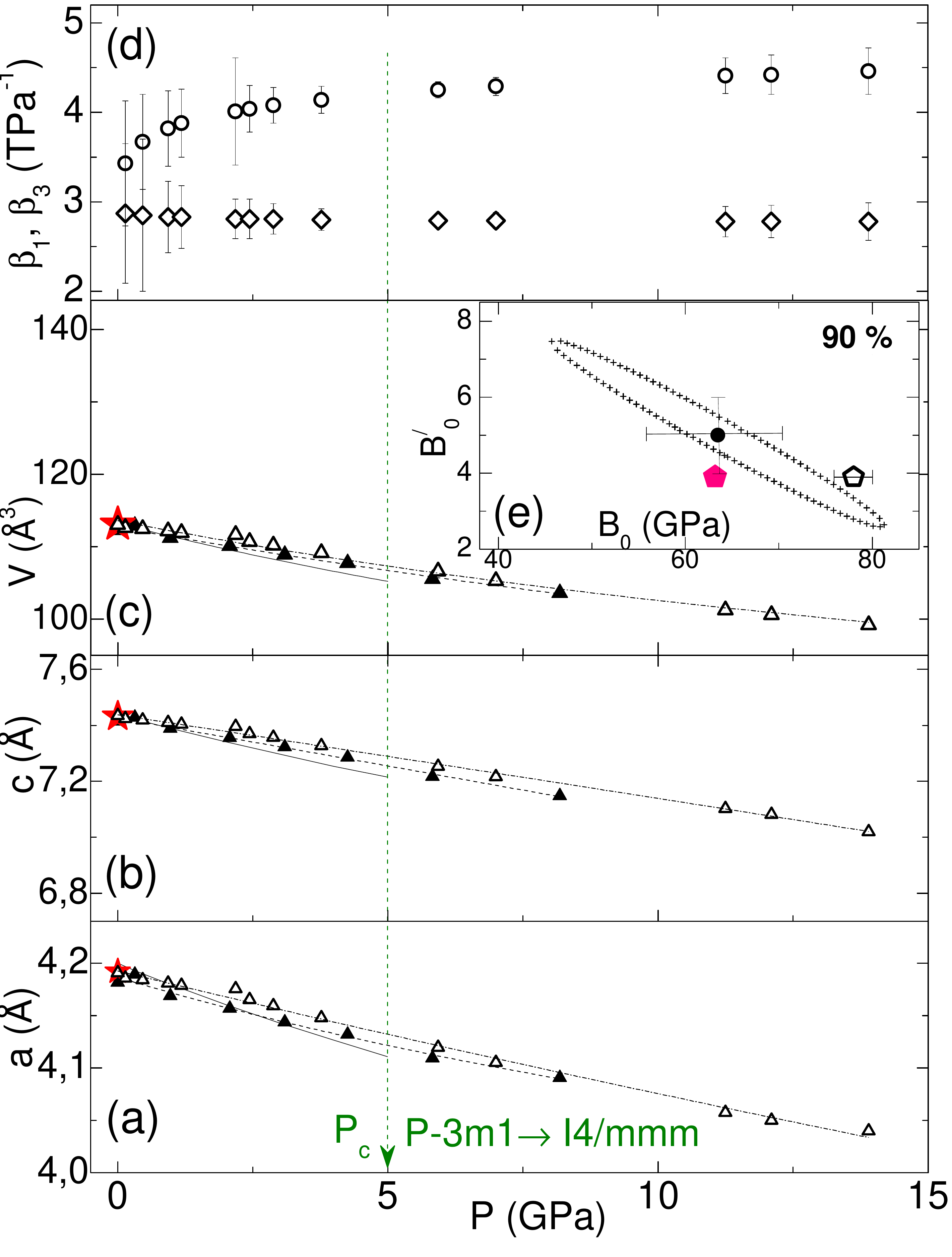}%
\caption{Baric evolution of  room-temperature lattice parameters [$a$, $c$, $V$, respectively (a), (b) and (c)] of $\alpha$ SrAl$_2$Si$_2$. \textit{Symbols}: as obtained from the Rietveld analysis of Fig.~\ref{Fig6-Sr-Diffractogram-2nd-run}. The lattice parameters of the two runs are shown as closed (first run) and open (second run) triangle symbols. \textit{Red star symbol} represents the values after decompression. \textit{Black solid line} represents the calculated lattice constants (no fitting, see text). \textit{Thin dashed and dash-dot black lines} represent the fit parameters of the first and second runs, respectively, to  Eq.~\ref{Eq-V-vs-P-Murnaghan} with the obtained B$_{0}$ and B$^\prime_{0}$ as shown in Table~\ref{Table2-Bulk-Pc-Expt}. The vertical green arrow denotes $P_c^{cal}$. 
(d) Baric evolution of principal-axis compressibilities 
as obtained from the experimental data of panels (a,b) of second run; we used the online software PASCal \cite{cliffe12-PASCal-compressibility-software}. 
The directions of the principal axes and their corresponding compressibilities with respect to the crystallographic axis are (0,0,1) for $\beta_{1}$ and (-0.942,0.334,0), (0.587,0.809,0) for $\beta_{2}$, $\beta_{3}$ respectively.  
({\it Inset (e)}) The 90$\%$ confidence ellipse \cite{gonzalez16-eosfit7}, depicting the correlation between B$_{0}$ and B$^\prime_{0}$: being strongly elongated with a negative slope indicates that an increase in one leads to a reduction in the other and vice versa. Also drawn are the individual error bars for B$_{0}$ and B$^\prime _{0}$ as obtained from their variance. \textit{Pink closed pentagon symbol} represents the theoretical B$_{0}$ and B$^\prime_{0}$ values of Table~\ref{Table1-ElasticConstant}. \textit{Black open pentagon symbol} represents the bulk modulus value B$_{0}$=78(2) GPa after fitting with fixed B$^\prime _{0}$=3.9.}
\label{Fig7-Sr-Param-2nd-run}%
\end{figure}
%
\subsection{Experimental Results\label{Experimental-Results}}
\subsubsection{$P$-dependent structure of $\alpha$ SrAl$_2$Si$_2$ at 300~K \label{subSec.Sr-Results} }
Representative pressure-dependent diffractograms of $\alpha$ SrAl$_2$Si$_2$ within the $P < $14~GPa range are shown in Fig.~\ref{Fig6-Sr-Diffractogram-2nd-run}.  Measurements were conducted in two runs: the first up to 8~GPa while the second being performed so as to reach the 14~GPa range; both manifest a high degree of reproducibility within the common pressure range.
Except for a very weak contamination with SrAlSi (less than 1 wt $\%$),
the Rietveld refinement of the diffractograms indicates a single $P\overline{3}m1$ phase with 
lattice parameters as given in Figs.~\ref{Fig7-Sr-Param-2nd-run}(a-c) (see also Fig.~\ref{Fig1-Sr-Table-Trig-Tetra}). The diffractograms do not reveal any $P\bar{3}m1 \xrightarrow[]{HP,300K} I4/mmm$ transition, in apparent contradiction with theoretical prediction of $\S$\ref{SubSec-Theoretical-Results}. This contradiction is related to the presence of a separating high activation barrier which was not taken into consideration when assuming thermodynamic condition of equal free energies in $\S$\ref{SubSec-Theoretical-Results}. This same apparent discrepancy is evident in various other systems such as CaAl$_2$Si$_2$, \cite{Strikos20-CaAl2Si2-Metastability} and pressure-induced transformation ({\it diamond-cubic-phase}$~\xrightarrow{}\beta${\it -Sn-phase}) in bulk Si \cite{Mizushima94-Si-Stability-Pressure-Phase-Transition}. In all cases, the absence of 
transformation within the experimental conditions is related to a large energy barrier that hinders any reconstructive paths that promote atomic and bonding rearrangements \cite{Sharma96-Pressure-Induced-Amorphization}.

Figures~\ref{Fig7-Sr-Param-2nd-run}(a-c) reveal a monotonically-decreasing lattice parameters, a behavior that confirms, satisfactorily, the theoretical predictions (solid lines) with no adjustable parameters within $P \leq P_{c}^{cal}\approx$5~GPa range. 
In particular, when fitting the baric evolution of the measured volume to Eq.\ref{Eq-V-vs-P-Murnaghan}, one obtains B$_{0}$ and B$^\prime_{0}$ values shown in Table~\ref{Table2-Bulk-Pc-Expt}.
The agreement between these empirically-determined B$_{0}$ and B$^\prime_{0}$  and the theoretically-calculated ones (Table~\ref{Table2-Bulk-Pc-Expt}) is quite satisfactorily if one takes into consideration that the standard deviation of the fit are not small, that the fit is conducted within a restricted region that guarantees a satisfactory hydrostatic condition, and that there is a strong correlation among B$_{0}$ and B$^\prime_{0}$ (as shown in the confidence ellipse \cite{gonzalez16-eosfit7} in the Inset (e) of Fig.~\ref{Fig7-Sr-Param-2nd-run}).

Finally, Table~\ref{Table1-ElasticConstant} compares the empirically-determined compressibilities [$\frac{\delta(a/a_0)}{\delta P}$ and $\frac{\delta(c/c_0)}{\delta P}$ evaluated from Fig.~\ref{Fig7-Sr-Param-2nd-run}(a,b)] with the theoretically-calculated compressibilities (see $\S$~\ref{SuSubSec-T=0-Calculation} and Eqs.~\ref{Eq-Compressibilities-along-a},~\ref{Eq-Compressibilities-along-c}). Evidently, these values do confirm the above-mentioned structural anisotropy which is related to the layered character: magnitudes of compressibilities along the c-axis are higher than those along the a-axis. In addition, both theoretical and experimental values are very close to each other supporting the validity of our theoretical and empirical analyses.
%
\begin{figure}[htbp]
\centering
\includegraphics[scale=0.28,trim=0.01mm 0.2mm 0.01mm 0.1mm,clip] {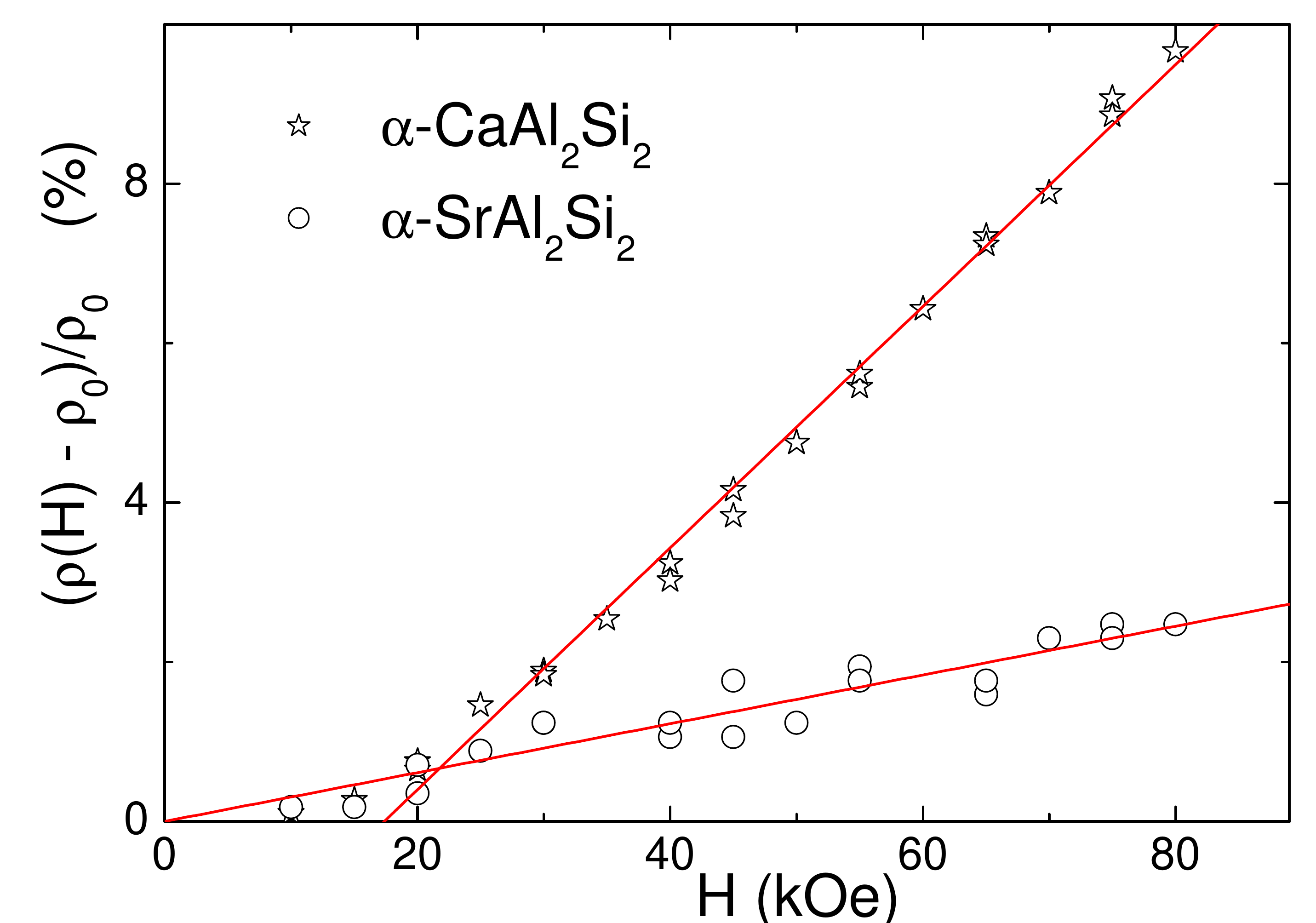}   
\caption{Magnetoresistivity of $\alpha$ SrAl$_2$Si$_2$ measured at 2.40(2)~K and ambient-pressure. Comparison with the magnetoresistivity of $\alpha$ CaAl$_2$Si$_2$, measured at the same conditions, emphasizes its weak linear-in-$H$ contribution (the linear slope of the Ca-based is 0.15$\%/$kOe while that of Sr-based is 0.03$\%/$kOe). That the fit of the magnetoresistivity of  $\alpha$ CaAl$_2$Si$_2$ does not pass through the origin is due to the fact that the linear magnetoresitivity is only manifested above a certain threshold field \cite{Costa18-CaAl2Si2-MagRes}.
}
\label{Fig8-SrAl2Si2-Ca-Norm-MagRes-vs-H}
\end{figure}
\subsubsection{Isothermal Magnetoresistvity of $ \alpha$ SrAl$_2$Si$_2$ \label{Sec.Results-LMR}}

We measured thermal evolution of both the magnetic susceptibility and electrical resistivity of $\alpha$ SrAl$_2$Si$_2$ and these do reproduce the corresponding curves reported in Refs.~\onlinecite{kauzlarich09-SrAl2SI2-structure-Thermoelectric,Lue11-Electronic-structure-(Sr-Y)Al2Si2, Zevalkink17-SrAl4-xSix-Making-Breaking-Bonds}. As an illustration, the resistivity, shown in Fig.~S2, does confirm the semimetallic character.\cite{Zevalkink17-SrAl4-xSix-Making-Breaking-Bonds} As mentioned above, the diamagnetic susceptibility and the semimetallic resistivity as well as the reported negative sign of the Seebeck coefficient of $\alpha$ SrAl$_2$Si$_2$ \cite{kauzlarich09-SrAl2SI2-structure-Thermoelectric,Lue11-Electronic-structure-(Sr-Y)Al2Si2} are consistent with the observation that the transport properties are dominated by the topology of the Fermi surfaces of hole and electron pockets. 

On considering a rigid-energy-band approximation for both $\alpha$ $A$Al$_2$Si$_2$ (Fig.~S3), one wonders whether the electrical properties (e.g. magnetoresistivity) of $\alpha$ SrAl$_2$Si$_2$ are similar to that of $\alpha$ CaAl$_2$Si$_2$  \cite{Costa18-CaAl2Si2-MagRes}. 
Fig.~\ref{Fig8-SrAl2Si2-Ca-Norm-MagRes-vs-H} shows that the isothermal field-dependent normalized magnetoresistivity of $\alpha$ SrAl$_2$Si$_2$, 
measured at 2.4~K. Evidently, it is neither saturating nor quadratic-in-$H$; rather, as compared to the reported linear-in-$H$ magnetoresistivity of $\alpha$ CaAl$_2$Si$_2$, it exhibits a weak linear-in-$H$ behavior, with the linear slope being almost 20~$\%$ of the value reported for its isomorph \cite{Costa18-CaAl2Si2-MagRes}. This is attributed to the reduced contribution of the $h_3$ pocket which, according to  Ref.~\onlinecite{Costa18-CaAl2Si2-MagRes}, is the one which fulfills Abrikosov's field and temperature conditions \cite{abrikosov1999quantum} required for the manifestation of linear magnetoresistivity.
%
\begin{figure*}[tbp] 
\centering
\noindent\includegraphics[scale=0.35]{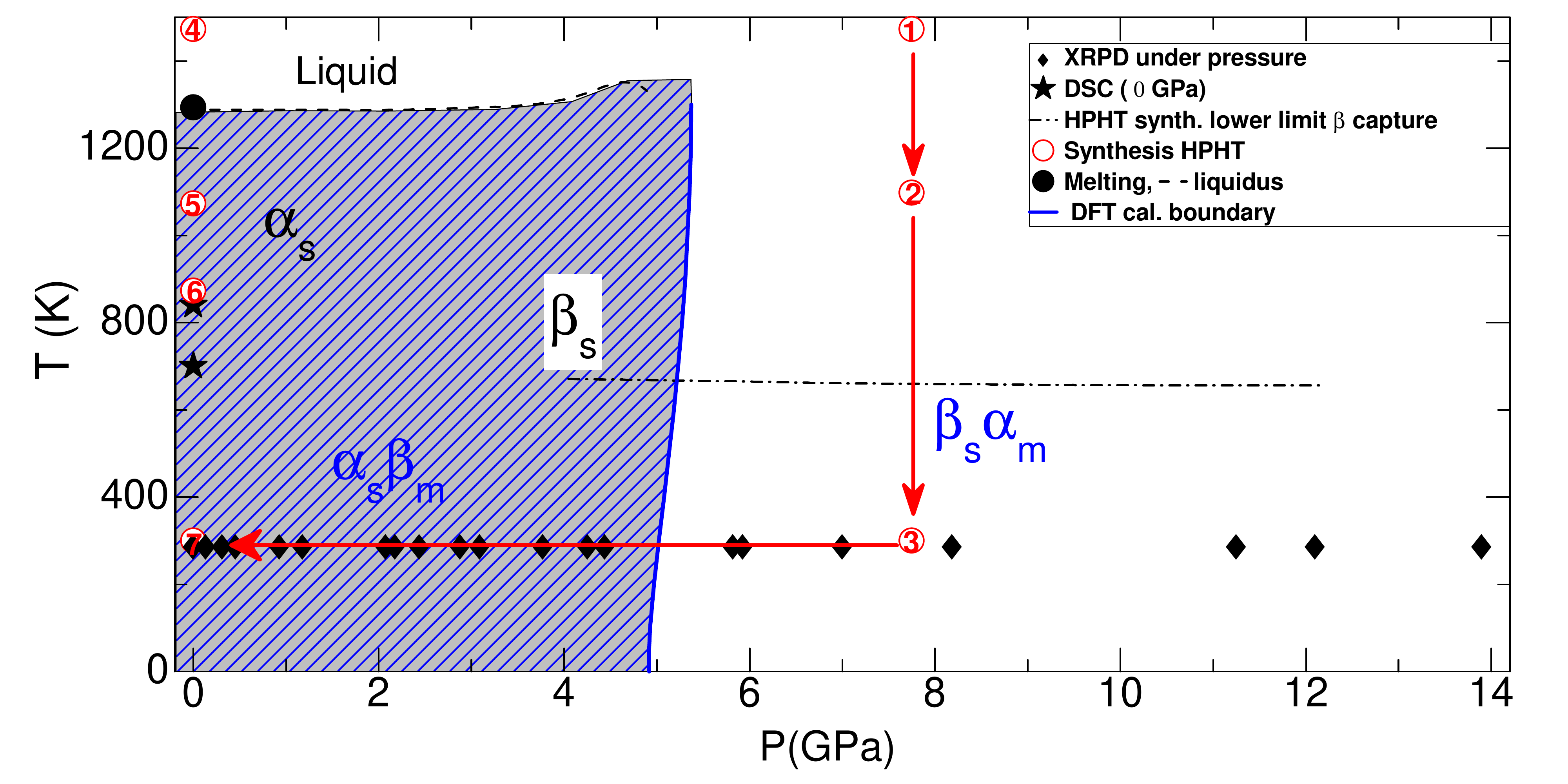}%
\caption{The equilibrium and nonequilibrium $P\text{\textendash} T$ diagram of SrAl$_2$Si$_2$.  The calculated equilibrium phase diagram, with boundaries marked by the blue line, is superimposed on the top of the diagram representing various experimental events as well as the extrapolated boundary lines. Information were collected from the synthesis data \cite{Zevalkink17-SrAl4-xSix-Making-Breaking-Bonds,Zevalkink20-SrAl2Si2-PrivateCom}, the ambient-pressure thermal analysis \cite{Zevalkink17-SrAl4-xSix-Making-Breaking-Bonds}, the ambient-pressure  melting point \cite{kauzlarich09-SrAl2SI2-structure-Thermoelectric}, the XRPD analysis (Ref.~9
, this work),  and the DFT-based calculations (this work). As discussed in the text, the metastability region lies to the right of the blue hatched area. To identify and emphasize the meta/stable character, the corresponding phases are indicated as $\alpha_{i}\beta_{j}$ where i,j $\equiv$ s (stable) or m (metastable). The liquidus is represented by the upper, almost horizontal curve, while the dash-dot curve indicates the range above which the $\beta$ phase can be quenched-and-captured; both curves are extrapolations that serve as visual guides. The long thick red arrows depict the path of the HPHT synthesis processes ($1\rightarrow2\rightarrow3\rightarrow7$). Ambient-pressure arc-melting/thermal treatment is indicated by the path $4\rightarrow5\rightarrow6\rightarrow7$. Depending on the selected preparation route, some steps may be skipped.
It is worth adding that a construction of a similar $P\text{\textendash} T$ diagram of CaAl$_2$Si$_2$ was hampered by its HPHT chemistry \cite{Tanaka13-CaAl2Si2-HPHT-Synthesis,Tanaka13-HPHT-Ca2Al3Si4-Synthesis,Strikos20-CaAl2Si2-Metastability}. 
}
 \label{Fig9-Sr-P-T-Phase-Diagram-Proposed}%
\end{figure*}
%
\subsubsection{Superposition of Empirical and Calculated $P\text{\textendash} T$  Diagrams \label{SubSubSec.Phase-Diagram}}

The calculated $P\text{\textendash} T$ diagram, delineated by the solid blue line in Fig.~\ref{Fig9-Sr-P-T-Phase-Diagram-Proposed}, is an equilibrium diagram which, as seen in $\S$\ref{subSec.Sr-Results}, is in an apparent disagreement with experiment. As we mentioned above, the thermodynamic-equilibrium analysis of $\S$~\ref{SubSec-Theoretical-Results} did not consider the activation barrier and as such this analysis is incapable of describing any nonequilibrium transformation or metastability. Accordingly, for constructing a diagram of transformations and for analyzing the synthesis and transformation routes, one should exercise particular care since different experimental {\it P-T} paths would not connect similar end-point polymorphic structural phases. \cite{Brazhkin06-Metastability-Transitions-PhaseDiagram,Parija18-Metastable-Phase-Acessability-Utility}
We believe that the $\beta$ phase is metastable and that its formation requires a much higher energy than that available at room temperature. Its formation requires the targeting of those {\it P-T} conditions whereat it is thermodynamically stable and afterwards follow the path wherein it can be kinetically trapped. 
Below we recall some synthesis procedures, involving different {\it P-T} trajectories, for the synthesis of the $\alpha$ and $\beta$ phases: this allows the delineation of the phase boundaries.

The construction of the diagram of nonequilibrium transformations, Fig.\ref{Fig9-Sr-P-T-Phase-Diagram-Proposed}, is based on the following reported events (namely, the HPHT synthesis events and differential scanning calorimetry analysis) \cite{kauzlarich09-SrAl2SI2-structure-Thermoelectric,Zevalkink17-SrAl4-xSix-Making-Breaking-Bonds,Zevalkink20-SrAl2Si2-PrivateCom}: 
(i) Subjecting an initially room-temperature global-energy-minimum $\alpha$ phase to a HPHT treatment 
(400$ \leq T\leq $1250 °C, 4$ \leq P\leq $9.5~GPa) followed by quenching under the same clamping pressure, the tetragonal $\beta$ SrAl$_2$Si$_2$ is energetically trapped (the $7\rightarrow3\rightarrow1\rightarrow2\rightarrow3\rightarrow7$ route). 

(ii)  Subjecting an initially room-temperature-stabilized, metastable $\beta$ phase to heating above 700~K under ambient-pressure and subsequent temperature ramp-down, the $\beta$ phase is transformed, irreversibly, back into the $\alpha$ phase (the $7\rightarrow6\rightarrow7$ route).

These two observations can also be symbolically expressed as
\begin{subequations}
\begin{flalign}
& \alpha \xrightarrow [irreversible] {HPHT} \beta,   \label{Eq-alpha-to-beta}\\
& \beta \xrightarrow [irreversible]{P_{a},HT} \alpha. \label{Eq-beta-to-alpha}
\end{flalign}
\end{subequations}
The  metastability character of SrAl$_2$Si$_2$, highlighted in Fig.~\ref{Fig9-Sr-P-T-Phase-Diagram-Proposed}, is a dominant feature within the studied $P\text{\textendash} T$ diagram: within the low-pressure blue-hatched region, the $\alpha$ phase is stable while the $\beta$ phase is metastable (hence the label $\alpha_s\beta_m$). In contrast, within the high-pressure, white region, the stability is expected to be reversed (hence the label $\beta_s\alpha_m$). 

The synthesis, the accessibility, the stabilization, and the functionality of each individual phase of SrAl$_2$Si$_2$ can be finely tuned by adjusting the control parameters (e.g. pressure, temperature, doping, and stoichiometry).\cite{Parija18-Metastable-Phase-Acessability-Utility} In essence this amounts to a selection of a trajectory within the configuration space that leads to a specified local free-energy minimum (if one intents to trap a specific structure under ambient pressure and temperature, without being annealed into another structure). 

The presence of the relatively high activation energy \big[separating e.g. the $\alpha$ and $\beta$ phases, see Refs.~\onlinecite{Turnbull81-Metastable-Structures-Metallurgy, Brazhkin06-Metastability-Transitions-PhaseDiagram,Recio15-High-Press-Sciences} and Figs.~\ref{Fig2-Srl2Si2-Barrier-Phases-E-V-H-P-curves}(a, b)\big]
marks a strong distinction among the various possible synthesis routes and, as a consequence, among the reversible/irreversible and equilibrium/nonequilibrium transformations, the different structural motifs and chemical bonding (Fig.~\ref{Fig1-Sr-Table-Trig-Tetra}), and different electronic structure (Fig.~S3).

Considering the metastable polymorphism of SrAl$_2$Si$_2$, it is no surprise that there are various emergent functionalities of this compound: the thermoelectricity \cite{kauzlarich09-SrAl2SI2-structure-Thermoelectric,zevalkink14-nonstoichiometry-Zintl-thermoelectric}, superconductivity \cite{Zevalkink17-SrAl4-xSix-Making-Breaking-Bonds}, distinct thermal character (Seebeck, thermal conductivity, \cite{kauzlarich09-SrAl2SI2-structure-Thermoelectric, Lue11-Electronic-structure-(Sr-Y)Al2Si2} and specific heat \cite{Zevalkink17-SrAl4-xSix-Making-Breaking-Bonds}), quantum linear magnetoresistivity (this work), and, possibly topological Dirac semimetallicity \cite{Su20-magnetotransport-topological-states-CaAl2Si2,Deng20-electronic-structure-Dirac-semimetal-CaAl2Si2}.

It is worth mentioning that nonequilibrim transformation that traps a metastable phase (e.g. $\alpha \xrightarrow[P\ge P_c,T\ge T_c]{hot~compression} \beta$, Eq.(\ref{Eq-alpha-to-beta}), afterwards rapid-quench under clamped-pressure) is a widely observed phenomenon.\cite{Turnbull81-Metastable-Structures-Metallurgy,ponyatovsky92-pressure-induced-amorphous-phases,Parija18-Metastable-Phase-Acessability-Utility}  
Generally such transformations lead to metastable structures: e.g.  perovskite-related structures \cite{Preparative-methods-solid-state-chem}, \ce{CaCO3} \cite{Bayarjargal18-CaCO3s-PhaseDiagram-DFT}, intermetallic disilicides/digermindes \ce{$AX$2}  ($A$= Ca,~Sr,~Ba;~$X$=Si,~Ge) \cite{evers77-pressure-MSi2,evers80-transformation-BaSi2,imai98-phase-transitions-BaSi2,Imai03-ASi2-PT-PhaseDiagram,Wang15-BaSi2-PhaseStability-Transitions,nishii07-amorphous-BaSi2}, rare-earth diantimonides $R$Sb$_2$  ($R$=Dy~-~Lu,Y),  \cite{merrill82-AB2-Type-Compounds-HT-HP}, \ce{HfO2},\cite{Parija18-Metastable-Phase-Acessability-Utility} \ce{V2O5},\cite{Parija18-Metastable-Phase-Acessability-Utility} and $A$Al$_2$Si$_2$ (Sr, Ba) \cite{Yamanaka04-BaAl2Si2,Zevalkink17-SrAl4-xSix-Making-Breaking-Bonds,Yamanaka04-BaAl2Si2,Narita08-BaAl2Si2-Hpressure-Raman}. 

\section{Discussion and Conclusion \label{Sec.Discussion-Conclusion}}
Generally, kinetic processes as those delineated in  Fig.\ref{Fig9-Sr-P-T-Phase-Diagram-Proposed} are influenced by various factors such as the energetics of the interfaces within the studied sample.\cite{Chen97-GrainSize-Metastability-Semiconductor-NanoCrystals} Accordingly, one expects that the  phase boundaries in the diagram of Fig.~\ref{Fig9-Sr-P-T-Phase-Diagram-Proposed} would depend on factors such as sample morphology and size, distribution of defects and strain, rate of heating, rate of applied pressure, and experimental time of observation. 
In fact, the exact value of e.g, the critical pressure of the $\alpha \xrightarrow{} \beta$ transition is expected to be bound by a lower and an upper limit \cite{Mizushima94-Si-Stability-Pressure-Phase-Transition}: the lower bound is determined by the barrier-less thermodynamics analysis while the upper bound is manifested in defect- and strain-free cases wherein energy barrier is maximum. The observed values, due to the finite concentration of defects (acting as centres of phase nucleation) and accumulated strain in real samples, lie somewhere in between these two limits \cite{Mizushima94-Si-Stability-Pressure-Phase-Transition,Chen97-GrainSize-Metastability-Semiconductor-NanoCrystals}.

Finally, it is tempting, in spite of the irreversibility/metastability character,  to employ the theoretical phase transformations of  Fig.~\ref{Fig9-Sr-P-T-Phase-Diagram-Proposed} so as to obtain a rough estimate of the latent heat 
$L$ involved in the $\alpha \xrightarrow{P_c, T_c} \beta$ transition of SrAl$_2$Si$_2$. 
For that purpose, let us assume that the Clausius–Clapeyron relation \cite{Blank13-Phase-Transformation-Book}
\begin{equation}
L={T\,\Delta V}\frac{dP}{dT} ,
\label{Eq-Clausius–Clapeyron}
\end{equation}
is valid across the calculated $P_c(T_c)$ phase boundary. Moreover, let us assume that L is similar to the one involved in the ambient-pressure high-temperature transition revealed in the DSC curves \cite{Zevalkink17-SrAl4-xSix-Making-Breaking-Bonds}, Eq.(\ref{Eq-beta-to-alpha}).
Then, on substituting in Eq.(\ref{Eq-Clausius–Clapeyron}) the experimental $T_c \approx$700~K and theoretical $ (\Delta V)_{T_c} \approx$ -0.035~cm$^{3}$/g and  $(\frac{dP}{dT})_{T_c} \approx$ 0.44~MPa/K, we obtain an exothermic $L_{cal} \approx$-11~J/g. It is assuring that the calculations do reproduce the correct sign (exothermic) of the reported transformation \cite{Zevalkink17-SrAl4-xSix-Making-Breaking-Bonds}. 

In summary, our combined theoretical and experimental studies on pressure-dependent structural properties of SrAl$_2$Si$_2$ highlight its metastability and (non)equilibrium  transformations. 
Our computational studies enabled the construction of the $P\text{\textendash} T$ phase diagram. However the ambient-temperature, high-pressure XRPD characterization does not follow the calculated equilibrium phase diagram. 
Based on the above analysis, we were able to reconcile the discrepancy between theory and experiment, to explain the stability/metastability of the $\alpha$ and $\beta$ phases, to justify the requirement of HPHT
treatment involving thermal shock to drive the transformation,
and to follow the subsequent influence of the structural rearrangement on the topology of the Fermi surfaces and how this is manifested in the evolution of the electric transport properties. 

As an outlook, let us recall that the diagram of Fig.\ref{Fig9-Sr-P-T-Phase-Diagram-Proposed}, being constructed out of  limited number of structural, synthesis-routes, and thermal analyses, is preliminary and incomplete. Nevertheless, being a generalization and summary of all currently available experimental and theoretical information, it  
serves the purposes of identifying the (meta)stability regions and reconciling the discrepancy between our predictions and experiments. 
It is hoped that this diagram in particular and this work in general would motivate further theoretical and experimental investigations on, e.g, different $P\text{\textendash} T$ paths extended over expanded regimes of HPHT experiments and detailed calculations of kinetic barriers. These further analyses would allow a better understanding of the correlation among the metastable structural motifs/trasformations, the electronic structure, the physico-chemical properties, and functionality of SrAl$_2$Si$_2$.

\section*{Supporting Information}
Baric  evolution  of  the  normalized area and normalized Full-Width at Half Maximum of  the  most  intense  Bragg peak of $\alpha$ SrAl$_2$Si$_2$, zero-field and ambient-pressure resistivity,   goodness of fit factors of Rietveld analysis of SrAl$_2$Si$_2$, and analysis of the electronic band structure and DOS of both $\alpha$ and $\beta$ phases of SrAl$_2$Si$_2$ (a comparison with SrAl$_2$Si$_2$ CaAl$_2$Si$_2$ is also given). 

\section*{Acknowledgments}
We are grateful to CNPq, FAPERJ and FAPEMIG for partial financial support, to UFV computation cluster at Universidade Federal de Viçosa, to H. S. Amorim for useful discussion,  and to Xpress beamline at Elettra Sincrotrone Trieste for allocating the beam time (Proposal 20185043). We are also grateful for partial support from project CALIPSOplus under Grant Agreement 730872 from the EU Framework Program for Research and Innovation HORIZON 2020. 

\onecolumn

\appendix
\section{Supporting Information}
\setcounter{figure}{0} 
\setcounter{table}{0}
\renewcommand{\theequation}{S\arabic{equation}}
\renewcommand{\thetable}{S\arabic{table}}   
\renewcommand{\thefigure}{S\arabic{figure}}
\renewcommand{\thesubsection}{.\Alph{subsection}}

\begin{table*}[htbp]
\scriptsize
\caption{Reliability factors $\chi^{2}$, $R_{p}$, $R_{wp}$ as obtained from Rietveld analysis of the diffractograms of $\alpha$-SrAl$_2$Si$_2$ (see Fig.6 of the main text). These were obtained with no correction for background.
More information on the chemical and structural parameters can be found in Figs.~1,6 of the main text.
}
\begin{tabular}{|c|cccccccccccccc|}
\hline
P (GPa) & 0 & 0.14 & 0.46 & 0.93 & 1.18  & 2.18  & 2.44 & 2.88 & 3.77 & 5.93 & 7.00 & 11.25 & 12.10 & 13.90 \\
\hline
$\chi^{2}$& 1.41 &2.58  &2.59  &3.02  &3.21  &3.46  &3.14  &4.14  &4.43  &3.74  &3.64  &3.51  &3.85  &4.35 \\
$R_{p}$   & 1.44 &1.58  &1.60  &1.83  &1.92  &2.03  &2.06  &2.14  &2.27  &2.06  &2.04  &1.98  &2.04  &2.20 \\
$R_{wp}$  & 2.36 &2.70  &2.65  &2.84  &2.94  &3.06  &2.98  &2.98  &3.08  &2.82  &2.79  &2.79  &2.88  &3.10 \\
\hline
\end{tabular}%
\end{table*}

\begin{figure}[htbp]
\centering
\includegraphics[scale=0.32,trim=0.01mm 0.2mm 0.01mm 0.1mm,clip] {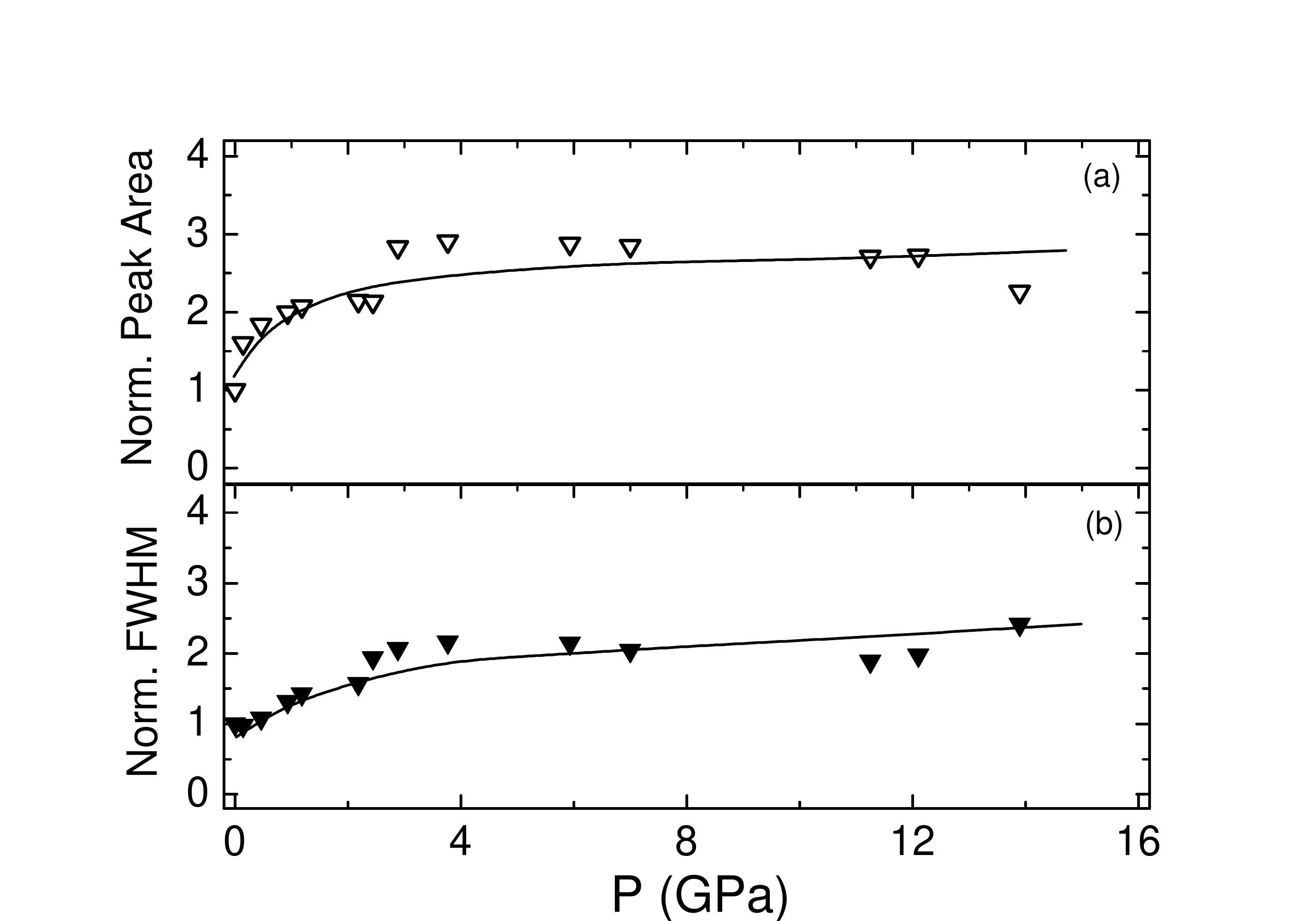}   
\caption{Baric evolution of (a) the normalized area and (b) normalized Full-Width at Half Maximum (FWHM) of the most intense Bragg, (011)/(101), peak of $\alpha$-SrAl$_2$Si$_2$ (see Fig.6 of the main text). Black lines are guides to the eye.}
\label{Fig10-SrAl2Si2-Sr-FWHM-vrs-pressure}
\end{figure}

\begin{figure}[htbp]
\centering
\includegraphics[scale=0.28,trim=0.1cm 0.1cm 0.1cm 0.1cm,clip] {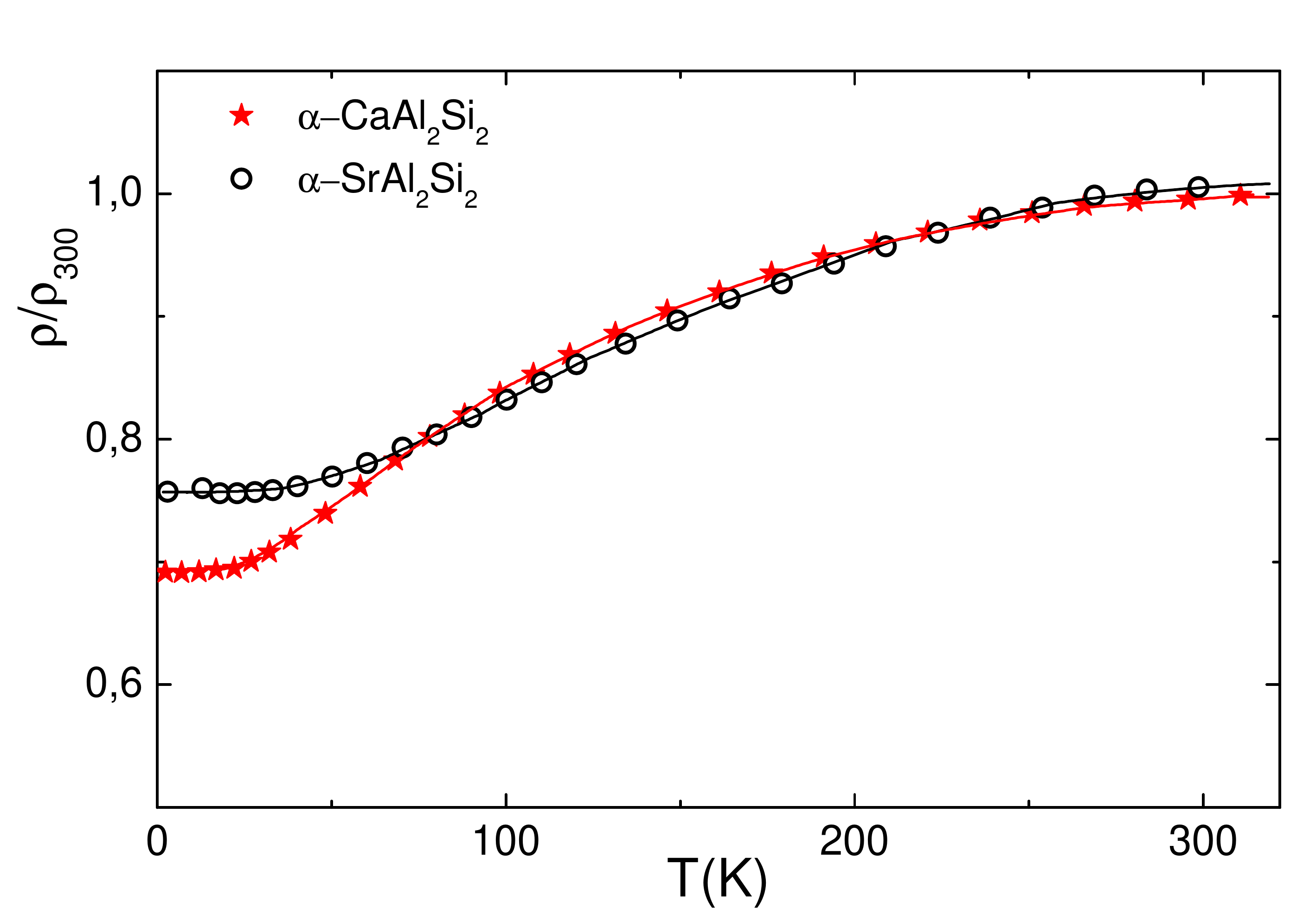}   
\caption{Thermal evolution of the normalized resistivity of $\alpha$-SrAl$_2$Si$_2$ and $\alpha$-CaAl$_2$Si$_2$ measured at ambient pressure and zero magnetic field. Both reveal characterisitc semimetallic behavior~\cite{kauzlarich09-SrAl2SI2-structure-Thermoelectric,Lue11-Electronic-structure-(Sr-Y)Al2Si2}, though that of $\alpha$-SrAl$_2$Si$_2$ is relatively much weaker. A detailed discussion on the thermal evolution of the resistivity of these $\alpha-R$Al$_2$Si$_2$ ismorphs can be found in Ref.~13
. Black and red lines are guides to the eye.
}
\label{Fig8-SrAl2Si2-Ca-Norm-MagRes-vs-H-updated}
\end{figure}

\begin{figure*}[ptb] 
\centering
\includegraphics[scale=0.60,trim=0.01mm 0.01mm 0.05mm 1.8mm,clip]{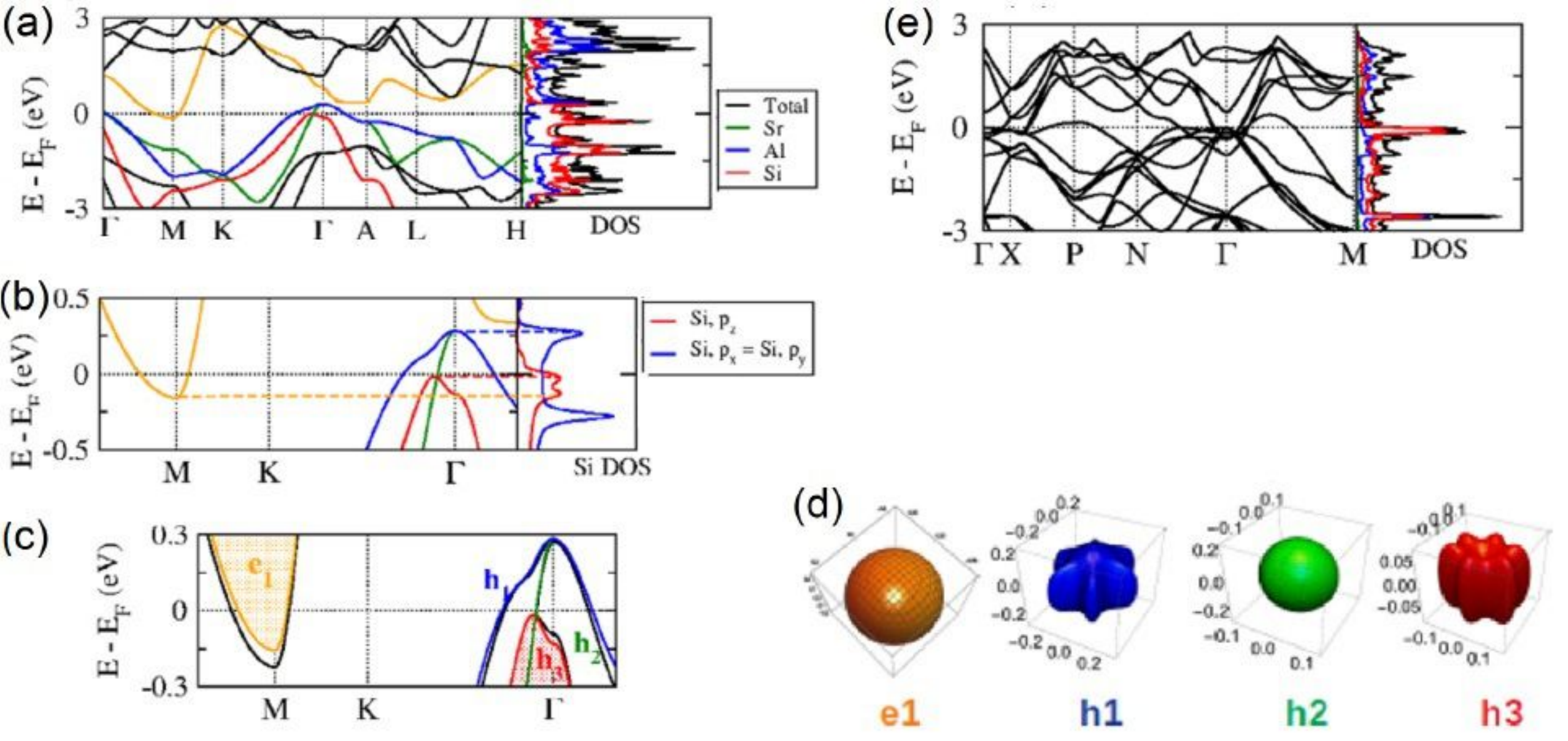}%
\caption{Electronic band structures of SrAl$_2$Si$_2$. (a) Calculated electronic structure and total DOS of $\alpha$ SrAl$_2$Si$_2$. Near the Fermi level, the DOS is dominated by Si orbitals. (b) Expanded view of the electron pocket (e$_1$, orange) and the three holes [h$_1$ (blue), h$_2$ (green), h$_3$ (red) pockets] along $ \Gamma-K-M$ direction.  The DOS is projected on the $p$ atomic orbitals of Si. (c) Comparison of the bands structure of $\alpha$ SrAl$_2$Si$_2$ (in color) and $\alpha$ CaAl$_2$Si$_2$ (in black).  (d) The Fermi surfaces of the
contributing pockets: the larger-sized and almost spherical $e_1$, $h_1$, $h_2$ and (the almost 25\% minor-in-height) $h_3$. These are very similar to the corresponding Fermi surfaces of CaAl$_2$Si$_2$ \cite{Costa18-CaAl2Si2-MagRes,Su20-magnetotransport-topological-states-CaAl2Si2}.  (e) Calculated electronic bands structure and total DOS of $\beta$ SrAl$_2$Si$_2$. 
All structures were relaxed before the commencement of the electronic structure calculation.}
\label{Fig5-SrAl2Si2-Ca-bands-DOS-Pockets}%
\end{figure*}

\newpage
\section{Electronic band structure and DOS analysis}

DFT-based electronic band structure and total DOS of the $\alpha$ and $\beta$ phases are shown in Figs.~\ref{Fig5-SrAl2Si2-Ca-bands-DOS-Pockets}(a~-~d). These diagrams, in good agreement with Refs.~\onlinecite{Zevalkink17-SrAl4-xSix-Making-Breaking-Bonds,Lue11-Electronic-structure-(Sr-Y)Al2Si2}, have direct relevance to the interpretation of the reported resistivity, Seebeck effect, thermal conductivity, thermoelectricity, and specific heat of these phases \cite{kauzlarich09-SrAl2SI2-structure-Thermoelectric,Lue11-Electronic-structure-(Sr-Y)Al2Si2,Zevalkink17-SrAl4-xSix-Making-Breaking-Bonds}. The following three remarks illustrate such a relevance: (i) The increase in $N_t(E_F)$  of the $\beta$ phase [Fig.~\ref{Fig5-SrAl2Si2-Ca-bands-DOS-Pockets}(e)] is consistent with its superconducting properties \cite{Zevalkink17-SrAl4-xSix-Making-Breaking-Bonds}. (ii) The configuration of multiple hole and electron pockets in the $\alpha$ phases is compatible with their semimetalicity being due to contributions from both the n-type and p-type carriers. (iii) This mixed contribution determines the strength of the magnetoresistivity as well as the sign and magnitude of both Hall and Seebeck effects. 

For illustrating the last remark, let us compare the electronic structure of $\alpha$ SrAl$_2$Si$_2$ with that of isomorphous $\alpha$ CaAl$_2$Si$_2$. Figs.~\ref{Fig5-SrAl2Si2-Ca-bands-DOS-Pockets}(b-c) exhibit, within the neighborhood of $M$ and $\Gamma$ points of both isomorphs, one electron pocket and three hole pockets; 
as evident, the total chemical substitution of Ca by the heavier and larger Sr (which amounts to a  chemically-induced negative pressure) leads to some, though relatively weak, change in the configuration of hole and electron pockets of $\alpha$ SrAl$_{2}$Si$_{2}$. On the one hand, there is an upward shift of the conduction band minimum which is assumed to be due to an induced chemical pressure effect rather than to a purely, direct, chemical substitution: the partial density of states, pDOS, Figs.~\ref{Fig5-SrAl2Si2-Ca-bands-DOS-Pockets}(a,b), indicates that, among the bands around the Fermi level, the major contributions is from Si orbitals rather than from  Ca/Sr related ones \cite{Peng18-AM2X2-Chemistry-Thermoelectric}. 
There is also, on the other hand, a corresponding chemically-induced variation in the hole pockets (see Figs.~\ref{Fig5-SrAl2Si2-Ca-bands-DOS-Pockets}(a,b)).
As the free charge carriers in intrinsic compensated semimetal are expected to be equal, then a decrease (due to upwards shift relative to $\alpha$ CaAl$_{2}$Si$_{2}$) in the electron-pocket contribution would be accompanied by a decrease in the hole-pockets contribution. A better visualization can be gained if we consider the angular momentum projected density of states of Si atoms. This suggests that $h_1$ and $h_2$ hole pockets (both derived from p$_x$ and p$_y$ orbitals) are hardly influenced while both $h_3$ and $e_1$ pockets (both derived  from p$_z$ orbital) do exhibit a rearrangement, most dominantly anti-parallel shift. This reduced contribution of $h_3$ pocket, we argue in $\S$3.2.2 of the main text, is the main reason behind the strong reduction in the linear-in-$H$ magnetoresistivity of $\alpha$ SrAl$_{2}$Si$_{2}$.

Finally, it is recalled that the sign of the Seebeck coefficient of $\alpha$ SrAl$_2$Si$_2$ is negative within the whole measured temperature range
\cite{kauzlarich09-SrAl2SI2-structure-Thermoelectric,Lue11-Electronic-structure-(Sr-Y)Al2Si2}. A similar Seebeck feature is evident in $\alpha$ CaAl$_2$Si$_2$  \cite{Kuo07-CaAl2Si2-Transport}.
Considering the above-mentioned compensated semimetalicity, the sign and magnitude of Seebeck coefficient must be related to the topology of the Fermi surface \cite{Markov19-Thermoelectric-Semimetals}. Fig.~\ref{Fig5-SrAl2Si2-Ca-bands-DOS-Pockets} shows the complexity of the Fermi surfaces of these pockets and, in addition, the large deviations from the simple parabolic band model. In spite  of these complications, some insight can be gained if we adopt the following constant-relaxation-time approximation of the intrinsic bipolar Seebeck coefficient \cite{Markov19-Thermoelectric-Semimetals,Mahan98-Good-Thermoelectrics}:    
\begin{equation}
S= -\frac{k_B}{2q} \Big[\frac{\sigma_e -\sigma_h}{\sigma_e +\sigma_h} (\frac{\epsilon_c-\epsilon_v}{k_B T}  +5) +\frac{\epsilon_c +\epsilon_v - 2E_F}{k_B T} \Big]
\label{Eq-Seebeck-bipolar}
\end{equation} 
where $k_B$  is Boltzmann constant, $q$ is elementary charge,  $\sigma_e$ and $\sigma_h$ are hole and electron conductivity, $\epsilon_c$ is the bottom of $e_1$ and $\epsilon_v$ is the top of the contributing $h$ bands. Then, $S$ of Eq.(\ref{Eq-Seebeck-bipolar}) must be dictated by the conductivity ratio ($\frac{\sigma_e}{\sigma_h}$, depending on, e.g., the asymmetry of the effective masses) and the relative position of $\epsilon_c$, $\epsilon_v$ and $E_F$. Figs.~\ref{Fig5-SrAl2Si2-Ca-bands-DOS-Pockets}(b,c) show that the involved energy separations are very  small, highlighting the decisive role of  $\frac{\sigma_e}{\sigma_h}$. Accordingly, we consider this close proximity of $\epsilon_c$, $\epsilon_v$ and $E_F$ together with the major contribution of the $e_1$ pocket (leading to $\sigma_e > \sigma_h$) to be the main factors behind the negative sign and moderate magnitude of $S(\text{T})$ of both $\alpha$ $A$Al$_2$Si$_2$ semimetals: for $A$=Sr, $-100<S(T<\text{300~K}) < 0 $~$\mu$V/K \cite{kauzlarich09-SrAl2SI2-structure-Thermoelectric,Lue11-Electronic-structure-(Sr-Y)Al2Si2} while for $A$=Ca, 
$-55 <S(T<\text{ 300~K}) < 0$~$\mu$V/K \cite{Kuo07-CaAl2Si2-Transport}.

\newpage

\textbf{Synopsis:} Using theoretical (free-energy, vibrational ), empirical (X-ray powder diffraction), and reported (synthesis and thermal) analyses, we construct {\it P-T} phase diagram of transformations for SrAl$_2$Si$_2$. 
There, semimetallic $\alpha$ SrAl$_2$Si$_2$ phase is separated from superconducting $\beta$ SrAl$_2$Si$_2$ phase by a high activation barrier that can be surpassed only under high-P and high-T conditions. 
Indeed, although equilibrium free-energy calculation predict a $\alpha \xrightarrow {\text{5GPa}} \beta$ transition, room-T, P-dependent diffractogram show no structural transformation up to 14~GPa.
\bibliography{SrAl2Si2-Metastability-archive}

\begin{mcitethebibliography}{61}
\bibitem[Not()]{Note-Four-Possible-phases}
In fact we carried out the comparative analysis for the following four possible
  phases \cite{Strikos20-CaAl2Si2-Metastability}: orthorhombic $Pnma$
  \cite{Condron07-BaAl2Si2,nolas14-physics-inorganic-clathrates}, orthorhombic
  $Cmcm$ \cite{Yamanaka04-BaAl2Si2,Carrillo-Cabrera02-EuGa2-xGe4+x}, tetragonal
  $I4/mmm$ \cite{Yamanaka04-BaAl2Si2}, and trigonal $P\bar{3}m1$
  \cite{Gladyshevsky67-Cryst-Str} phases. For brevity we limit our description
  to only the two low-lying phases which were observed experimentally.\relax
\mciteBstWouldAddEndPunctfalse
\mciteSetBstMidEndSepPunct{\mcitedefaultmidpunct}
{}{\mcitedefaultseppunct}\relax
\EndOfBibitem
\end{mcitethebibliography}
\end{document}